\documentclass[twocolumn]{aastex631}

\newcommand{\sersic}{{Sérsic}\xspace}%

\usepackage{bm}
\usepackage{multirow}
\usepackage{comment}
\usepackage{xspace}
\usepackage{amsmath} 
\usepackage{relsize,type1cm}
\usepackage{threeparttable}

\begin{document}

\title{The $M_{\rm BH}-M_*$ relation up to $z\sim2$ through decomposition of COSMOS-Web NIRCam images}

\correspondingauthor{Takumi S. Tanaka}
\email{takumi.tanaka@ipmu.jp}

\author[0009-0003-4742-7060]{Takumi S. Tanaka}
\affiliation{Department of Astronomy, Graduate School of Science, The University of Tokyo, 7-3-1 Hongo, Bunkyo-ku, Tokyo, 113-0033, Japan}
\affiliation{Kavli Institute for the Physics and Mathematics of the Universe (WPI), The University of Tokyo Institutes for Advanced Study, The University of Tokyo, Kashiwa, Chiba 277-8583, Japan}
\affiliation{Center for Data-Driven Discovery, Kavli IPMU (WPI), UTIAS, The University of Tokyo, Kashiwa, Chiba 277-8583, Japan}

\author[0000-0002-0000-6977]{John D. Silverman}
\affiliation{Department of Astronomy, Graduate School of Science, The University of Tokyo, 7-3-1 Hongo, Bunkyo-ku, Tokyo, 113-0033, Japan}
\affiliation{Kavli Institute for the Physics and Mathematics of the Universe (WPI), The University of Tokyo Institutes for Advanced Study, The University of Tokyo, Kashiwa, Chiba 277-8583, Japan}
\affiliation{Center for Data-Driven Discovery, Kavli IPMU (WPI), UTIAS, The University of Tokyo, Kashiwa, Chiba 277-8583, Japan}
\affiliation{Center for Astrophysical Sciences, Department of Physics \& Astronomy, Johns Hopkins University, Baltimore, MD 21218, USA}

\author[0000-0001-8917-2148]{Xuheng Ding}
\affiliation{Department of Astronomy, Graduate School of Science, The University of Tokyo, 7-3-1 Hongo, Bunkyo-ku, Tokyo, 113-0033, Japan}
\affiliation{Kavli Institute for the Physics and Mathematics of the Universe (WPI), The University of Tokyo Institutes for Advanced Study, The University of Tokyo, Kashiwa, Chiba 277-8583, Japan}
\affiliation{Center for Data-Driven Discovery, Kavli IPMU (WPI), UTIAS, The University of Tokyo, Kashiwa, Chiba 277-8583, Japan}

\author[0000-0003-3804-2137]{Knud Jahnke}
\affiliation{Max-Planck-Institut für Astronomie, Königstuhl 17, D-69117 Heidelberg, Germany}

\author[0000-0002-3683-7297]{Benny Trakhtenbrot}
\affiliation{School of Physics and Astronomy, Tel Aviv University, Tel Aviv 69978, Israel}

\author[0000-0003-3216-7190]{Erini Lambrides}
\altaffiliation{NPP Fellow.}
\affiliation{NASA-Goddard Space Flight Center, Code 662, Greenbelt, MD, 20771, USA}

\author[0000-0003-2984-6803]{Masafusa Onoue}
\affiliation{Kavli Institute for the Physics and Mathematics of the Universe (WPI), The University of Tokyo Institutes for Advanced Study, The University of Tokyo, Kashiwa, Chiba 277-8583, Japan}
\affiliation{Kavli Institute for Astronomy and Astrophysics, Peking University, Beijing 100871, China}

\author[0000-0001-6102-9526]{Irham Taufik Andika}
\affiliation{Technical University of Munich, TUM School of Natural Sciences, Department of Physics, James-Franck-Str. 1, D-85748 Garching, Germany}
\affiliation{Max-Planck-Institut f\"{u}r Astrophysik, Karl-Schwarzschild-Str. 1, D-85748 Garching, Germany}

\author[0000-0002-0101-6624]{Angela Bongiorno}
\affiliation{INAF–Osservatorio Astronomico di Roma, Via Frascati 33, I-00078, Monte Porzio Catone, Italy}

\author[0000-0002-9382-9832]{Andreas L. Faisst}
\affiliation{Caltech/IPAC, 1200 E. California Blvd. Pasadena, CA 91125, USA}

\author[0000-0001-9885-4589]{Steven Gillman}
\affiliation{Cosmic Dawn Center (DAWN), Denmark}
\affiliation{DTU Space, Technical University of Denmark, Elektrovej, Building 328, 2800, Kgs. Lyngby, Denmark}

\author[0000-0003-4073-3236]{Christopher C. Hayward}
\affiliation{Center for Computational Astrophysics, Flatiron Institute, 162 Fifth Avenue, New York, NY 10010, USA}

\author[0000-0002-3301-3321]{Michaela Hirschmann}
\affiliation{Institute for Physics, Laboratory for Galaxy Evolution and Spectral Modelling, EPFL, Observatoire de Sauverny, Chemin Pegasi 51, 1290 Versoix, Switzerland}

\author[0000-0002-6610-2048]{Anton Koekemoer}
\affiliation{Space Telescope Science Institute, 3700 San Martin Drive, Baltimore, MD 21218, USA}

\author[0000-0002-5588-9156]{Vasily Kokorev}
\affiliation{Kapteyn Astronomical Institute, University of Groningen, P.O. Box 800,
NL-9700AV Groningen, the Netherlands}

\author[0000-0002-9252-114X]{Zhaoxuan Liu}
\affiliation{Department of Astronomy, Graduate School of Science, The University of Tokyo, 7-3-1 Hongo, Bunkyo-ku, Tokyo, 113-0033, Japan}
\affiliation{Kavli Institute for the Physics and Mathematics of the Universe (WPI), The University of Tokyo Institutes for Advanced Study, The University of Tokyo, Kashiwa, Chiba 277-8583, Japan}
\affiliation{Center for Data-Driven Discovery, Kavli IPMU (WPI), UTIAS, The University of Tokyo, Kashiwa, Chiba 277-8583, Japan}

\author[0000-0002-4872-2294]{Georgios E. Magdis}
\affiliation{Cosmic Dawn Center (DAWN), Denmark}
\affiliation{DTU-Space, Technical University of Denmark, Elektrovej 327, DK2800 Kgs. Lyngby, Denmark}
\affiliation{Niels Bohr Institute, University of Copenhagen, Jagtvej 128, DK-2200 Copenhagen N, Denmark}

\author[0000-0002-7093-7355]{Alvio Renzini}
\affiliation{Osservatorio Astronomico di Padova, Vicolo dell’Osservatorio 5, Padova, I-35122, Italy}

\author[0000-0002-0930-6466]{Caitlin Casey}
\affiliation{Cosmic Dawn Center (DAWN), Denmark}
\affiliation{The University of Texas at Austin, 2515 Speedway Boulevard Stop C1400, Austin, TX 78712, USA}

\author[0000-0003-4761-2197]{Nicole E. Drakos}
\affiliation{Department of Physics and Astronomy, University of Hawaii, Hilo, 200 W Kawili St, Hilo, HI 96720, USA}

\author[0000-0002-3560-8599]{Maximilien Franco}
\affiliation{The University of Texas at Austin, 2515 Speedway Boulevard Stop C1400, Austin, TX 78712, USA}

\author[0000-0002-0236-919X]{Ghassem Gozaliasl}
\affiliation{Department of Computer Science, Aalto University, PO Box 15400, Espoo, FI-00 076, Finland}
\affiliation{Department of Physics, Faculty of Science, University of Helsinki, 00014-Helsinki, Finland}

\author[0000-0001-9187-3605]{Jeyhan Kartaltepe}
\affiliation{Laboratory for Multiwavelength Astrophysics, School of Physics and Astronomy, Rochester Institute of Technology, 84 Lomb Memorial Drive, Rochester, NY 14623, USA}

\author[0000-0001-9773-7479]{Daizhong Liu}
\affiliation{Max-Planck-Institut für extraterrestrische Physik, Gießenbachstraße 1, 85748 Garching b. München, Germany}

\author[0000-0002-9489-7765]{Henry Joy McCracken}
\affiliation{Institut d’Astrophysique de Paris, UMR 7095, CNRS, and Sorbonne Universit´e, 98 bis boulevard Arago, F-75014 Paris, France}

\author[0000-0002-4485-8549]{Jason Rhodes}
\affiliation{Jet Propulsion Laboratory, California Institute of Technology, 4800 Oak Grove Drive, Pasadena, CA 91001, USA}

\author[0000-0002-4271-0364]{Brant Robertson}
\affiliation{Department of Astronomy and Astrophysics, University of California, Santa Cruz, 1156 High Street, Santa Cruz, CA 95064, USA}

\author[0000-0003-3631-7176]{Sune Toft}
\affiliation{Cosmic Dawn Center (DAWN), Denmark}
\affiliation{Niels Bohr Institute, University of Copenhagen, Jagtvej 128, DK-2200 Copenhagen N, Denmark}


\begin{abstract}
Our knowledge of relations between supermassive black holes and their host galaxies at $z\gtrsim1$ is still limited, even though being actively sought out to $z\sim6$.
Here, we use the high resolution and sensitivity of JWST to measure the host galaxy properties for 107 X-ray-selected type-I AGNs at $0.68<z<2.5$ with rest-frame optical/near-infrared imaging from COSMOS-Web and PRIMER.
Black hole masses ($\log\left(M_{\rm BH}/M_\odot\right)\sim6.9-9.6$) are available from previous spectroscopic campaigns.
We extract the host galaxy components from four NIRCam broadband images and the HST/ACS F814W image by applying a 2D image decomposition technique.
We detect the host galaxy for $\sim90\%$ of the sample after subtracting the unresolved AGN emission.
With host photometry free of AGN emission, we determine the stellar mass of the host galaxies to be $\log\left(M_*/M_\odot\right)\sim9.5-11.6$ through SED fitting and measure the evolution of the mass relation between SMBHs and their host galaxies.
Considering selection biases and measurement uncertainties, we find that the $M_\mathrm{ BH}/M_*$ ratio evolves as $\left(1+z\right)^{0.48_{-0.62}^{+0.31}}$ thus remains essentially constant or exhibits mild evolution up to $z\sim2.5$.
We also see an amount of scatter ($\sigma_{\mu}=0.30^{+0.14}_{-0.13}$), similar to the local relation and consistent with low-$z$ studies, and a non-causal cosmic assembly history where mergers contribute to the statistical averaging towards the local relation is still feasible.
We highlight improvements to come with larger samples from JWST and, particularly, Euclid, which will exceed the statistical power of current wide and deep surveys.
\end{abstract}



\section{Introduction} \label{sec:intro}

With our understanding that galaxies grow by increasing their stellar mass through mergers and in situ star formation from gas accretion, there are still many unresolved questions in galaxy evolution.
One of the most important challenges in galaxy formation is understanding the physical processes that relate the growth of supermassive black holes (SMBHs) alongside the growth of the galaxies that harbor them.
Observational studies, mainly in the local universe, have unveiled tight correlations between the mass of SMBHs ($M_{\rm BH}$) and the physical properties of their host galaxies, such as stellar velocity dispersion $\sigma_*$ and stellar mass $M_*$ \citep{Magorrian1998, Ferrarese2000, Marconi2003, Haring2004, Gultekin2009, Graham2011, Beifiori2012, Kormendy2013, Reines2015}.
How and through what physical processes such a tight relation is formed is still unclear, and the origin of the mass relation can shed light on the evolution of not only SMBHs but also galaxies.

A widely considered scenario, in answer to this question, is a co-evolution scheme, where galaxies and black holes mutually increase their mass at a correlated pace.
As a potential physical cause for a co-evolution scenario, some studies implement active galactic nuclei (AGN) feedback, where the energy released from AGNs heats the gas and controls star formation or gas accretion through radio jets or an AGN winds \citep[e.g.][]{Springel2005, DiMatteo2008, Hopkins2008, Fabian2012, DeGraf2015, Harrison2017}.
Additionally, studies support a common gas supply simultaneously fueling both SMBHs and their host galaxies by increasing the BH accretion rate and star formation rate  \citep[SFR,][]{Cen2015, Menci2016}.
On the other hand, others have shown that, even in the absence of a close physical connection between SMBHs and host galaxies, the mass relation can be achieved through a non-casual connection; major mergers have averaged the mass relation statistically \citep[cosmic averaging scenario;][]{Peng2007,Hirschmann2010,Jahnke2011}.

To unravel the cause of the mass relation, an effective approach is observing the relation between $M_{\rm BH}$ and $M_*$ throughout cosmic history.
With such observations, we can directly determine whether the relation in the local universe, both its ratio and dispersion, evolves with redshift.
Then, comparisons of the observational results with simulations \citep[e.g.,][]{Ding2020_simu, Habouzit2021, Ding2022_modelcomp} based on various physical models can allow us to discuss the physical processes that establish the galaxy-BH relations and further constrain the physics of black hole formation and galaxy evolution.

Before the advent of James Webb Space Telescope (JWST), statistical studies using 2D image decomposition analyses were conducted using images obtained by Hubble Space Telescope (HST) \citep[e.g.,][]{Peng2006,Jahnke2009,Bennert2011highz,Cisternas2011,Simmons2011,Simmons2012,Schramm2013,Mechtley2016,Ding2020_HST,Bennert2021,LiJ2023} and ground-based telescopes including Subaru's Hyper Suprime-Cam (HSC) \citep{Ishino2020,Li2021_HSC}, and Pan-STARRS1 (PS1) \citep{Zhuang2023nature}.
In summary, these studies have concluded that the relation between $M_{\rm BH}$ and $M_*$ does not evolve with redshift at $z\lesssim2$. 
However, studies using a large statistical and universal sample have yet to be achieved at $z>1$. At these redshifts, we can obtain information longer than the 4000~\AA\ break to constrain $M_*$ from observations at near-infrared wavelengths. 
However, there are a limited number of previous studies using near-infrared data at $z\gtrsim1$ \citep[e.g.,][utilizing HST/NICMOS or HST/WFC3]{Jahnke2009, Simmons2012, Mechtley2016, Ding2020_HST}.
Other studies carry out statistical AGN samples using spectral energy distribution (SED) fitting based 1D decomposition method \citep[e.g.,][]{Merloni2010, Sun2015, Suh2020}.

JWST is now revolutionizing the field of AGN - host galaxy relations up to $z\sim6$ and beyond based on its high spatial resolution and unprecedented sensitivity.
For instance, \cite{Ding2022_CEERS} applied a 2D decomposition analysis on early JWST NIRCam data from CEERS that successfully detected the host galaxies of five quasars at $z\sim 1.6-3.5$ from the SDSS DR17Q catalog \citep{Lyke2020}.
They also succeeded in detecting clear substructure and performed pixel-by-pixel SED fitting for one of the five targets, SDSSJ1420+5300A, at $z\sim1.6$.
Other studies have also performed 2D decompositions of AGN host galaxies using JWST imaging; for instance, \cite{Li2023} have analyzed a galaxy that is one of the most promising candidates for having a recoiling SMBH ($z\sim0.36$) while \cite{Kocevski2023} present the host properties of five X-ray-luminous AGNs ($3 < z < 5$) in CEERS.
\cite{Zhang2023} also utilized JWST NIRCam data to assess the validity of $M*$ estimated from 1D decomposition (spectrum-based) method for the HETDEX type-I AGNs ($2<z<2.5$). 

At $z>6$, \cite{Ding2022_z6} conducted decomposition analysis of two low-luminosity quasars, thus representing the highest-redshift record for detection of host stellar emission. 
They suggest that $z\sim6$ low luminosity quasars have a mass relation consistent with the local relation after considering selection biases and measurement uncertainties, albeit with a small sample. Equally remarkable, JWST studies of high-$z$ AGNs are revealing a higher abundance of lower mass black holes that are actively accreting within very dusty and compact galaxies \citep[e.g.,][]{Onoue2023, Kocevski2023, Matthee2023, Harikane2023, Maiolino2023, Greene2023, Kocevski2024, Wang2024, Akins2024}.
However, in general, the sample sizes of these high-$z$ AGNs and the accuracy of $M_*$ estimation are still limited.
Therefore, the redshift evolution of the mass relation remains highly uncertain.

Therefore, as an important next step to investigate the evolution of the mass relation, we perform a 2D decomposition analysis with JWST/NIRCam data for a sample whose redshift range significantly improves upon what was statistically analyzed before the JWST era, further bridging the gap between low-$z$ statistical and limited high-$z$ studies. 
We use a sample larger than previous studies using JWST that reaches up to $z\sim2-3$ when AGN and star formation activities peaked in cosmic history. 
With NIRCam images of $N=107$ AGNs in COSMOS-Web \citep{Casey2022} and PRIMER-COSMOS, we conduct 2D decomposition analyses and then statistically discuss the evolution of the $M_{BH}-M_*$ relation with consideration of the selection bias and measurement uncertainty \citep{Lauer2007,Shen2010,Schulze2011,Schulze2014}.
In addition, we report on the ability to accurately model the JWST PSF in each band and the impact on the derived host galaxy properties.

This paper is organized as follows.
Section~\ref{sec:data} describes the JWST/NIRCam data and the sample selection.
Section~\ref{sec:method} describes the detailed analysis method including 2D image decomposition and careful PSF modeling, SED fitting, and constrcution of mock data.
In Section~\ref{subsec:morphology}, we present our fitting results and discuss the PSF effect on the results.
Then, we show the evolution of $M_{\rm BH}-M_*$ relation in Section~\ref{subsec:mass-mass} with considering the selection bias. Also, we discuss the possibility of scatter evolution in the mass relation and summarize the challenges with 2D decomposition methods in Section~\ref{sec:discussion}.
We present the conclusion and prospects for future studies in Section~\ref{sec:conclusion}.
In this paper, all magnitude are AB magnitude \citep{OkeAB}, and we assume a standard cosmology with $H_0 = 70~{\rm km~s^{-1}~Mpc^{-1}}$, $\Omega_m = 0.30$, and $\Omega_\Lambda=0.70$.


\section{Data} \label{sec:data}
\subsection{COSMOS-Web}\label{subsec:data_cw}

\begin{figure}
\epsscale{1.1}
  \plotone{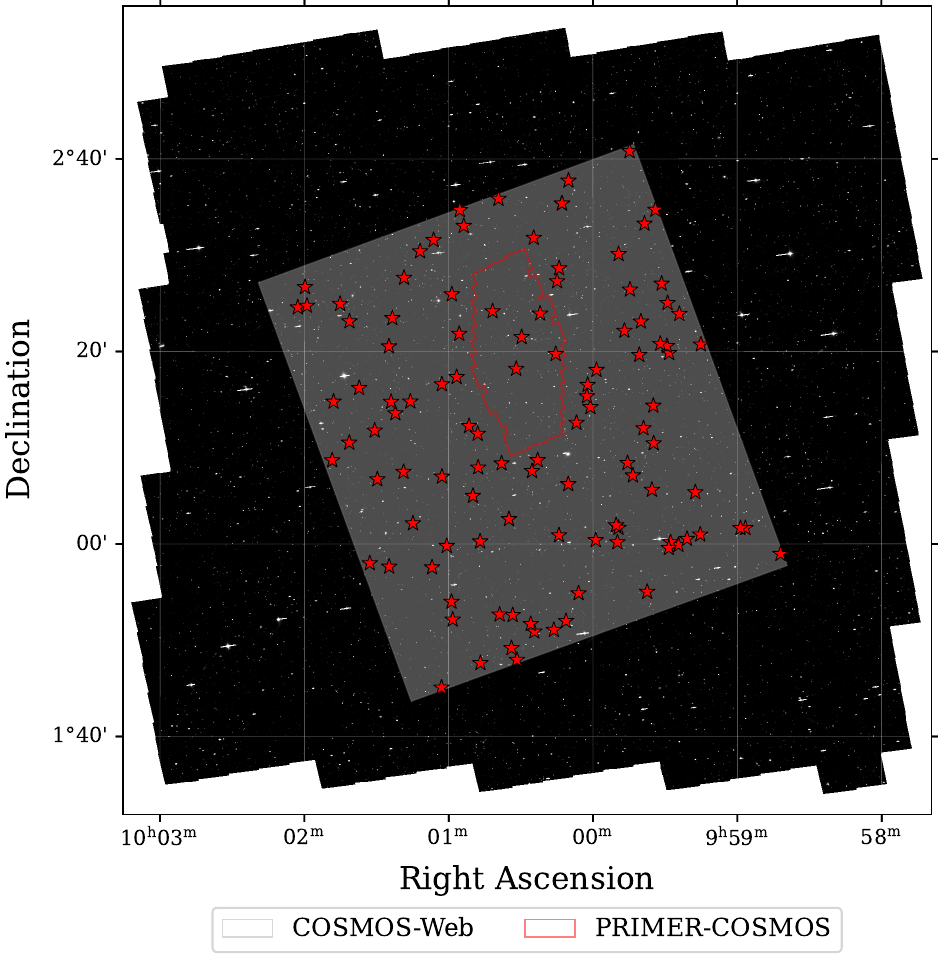}
  \caption{Location of the sample compared to the footprints of COSMOS-Web and PRIMER-COSMOS.
  The background is the mosaic image of HST/ACS F814W \citep{Koekemoer2007} for the COSMOS field.
  Red stars show the position of each AGN.
  The region enclosed by the solid red line shows the PRIMER-COSMOS field, and gray shaded region corresponds to the COSMOS-Web field.
  \label{fig:footprints}}
\end{figure}

COSMOS-Web \citep[PI: Jeyhan Kartaltepe and Caitlin Casey, GO1727, see][for the overview]{Casey2022} is a 270-hour treasury survey program in JWST Cycle~1, covering 0.54~${\rm deg}^2$ with NIRCam \citep{Rieke2023} in four filters (F115W, F150W, F277W, F444W) and 0.19 ${\rm deg}^2$ with MIRI \citep{Bouchet2015} using F770W.
Due to the large field, the COSMOS-Web field was split into twenty tiles.

The data are reduced with the JWST Calibration Pipeline\footnote{\url{https://github.com/spacetelescope/jwst}} \citep{jwst_pipeline} version 1.14.0 and the calibration Reference Data System version 1223.
The $5\sigma$ depth in an aperture with a radius of 0\farcs15 ranges from 26.7 to 27.5~mag in F115W and 27.5 to 28.2~mag in F444W, depending on the number of integrations \citep[also see Section~2.1 of][]{Casey2022}. 
The mosaic images have a resolution of 0\farcs030/pixel.
Details of the image reduction process will be described in Franco et al.,(in preparation).
In addition to NIRCam four-band data, we use HST/ACS F814W data \citep{Koekemoer2007}.
In this study, we do not use MIRI data because it is challenging to apply the 2D decomposition method (Section.~\ref{subsec:decomposition}) due to its lower spatial resolution and larger PSF.

\subsection{PRIMER}\label{subsec:data_primer}
Public Release IMaging for Extragalactic Research (PRIMER, PI: James Dunlop, GO1837) is a 195-hour treasury program of JWST Cycle~1, targeting two equatorial HST CANDELS Legacy Fields: COSMOS and UDS.
PRIMER-COSMOS covers 144 ${\rm arcmin}^2$ with eight NIRCam \citep{Rieke2023} filters (F090W, F115W, F150W, F200W, F277W, F356W, F410M, F444W) and 112 ${\rm arcmin}^2$ with two MIRI \citep{Bouchet2015} filters (F770W and F1800W) in the COSMOS field.
The processed COSMOS-PRIMER data consists of one mosaic image.
In this analysis, we use NIRCam eight-band data and HST/ACS F814W data.
The PRIMER-COSMOS data are reduced with the JWST Calibration Pipeline \citep{jwst_pipeline} version 1.8.3 and the calibration Reference Data System version 1017.
The $5\sigma$ depth in an aperture with a radius of 0\farcs15 has a wide range from 27.9 to 28.3~mag in F090W and 28.4 to 28.9~mag in F444W, depending on the number of integrations, $\sim$1~mag deeper than the COSMOS-Web data.
The mosaic images have a resolution of 0\farcs030/pixel.

\subsection{Broad-line AGN sample}\label{subsec:sample_selection}
To evaluate the relation between $M_*$ and $M_{\rm BH}$, we use the type-I AGN sample with $M_{\rm BH}$ estimates available in \cite{Schulze2015, Schulze2018}.
\cite{Schulze2015} presents the redshift evolution of AGN population based on spectroscopically observed type-I AGNs from zCOSMOS \citep{Lilly2007,Lilly2009}, VVDS \citep{LeFevre2005, LeFevre2013, Garilli2008}, and SDSS \citep{Schneider2010, Shen2012}. \cite{Schulze2018} provides the properties of X-ray selected and spectroscopically-confirmed type-I AGNs in the FMOS-COSMOS survey \citep{Kashino2013, Silverman2015}.
Here, we select the targets in \cite{Schulze2015} and \cite{Schulze2018} that are also detected by Chandra \citep[Chandra-COSMOS Survey;][]{Elvis2009, Civano2012} or XMM-Newton \citep[XMM-COSMOS;][]{Cappelluti2009, Brusa2010}.
The 2-10~keV flux sensitivity is $7.3\times10^{-16}~{\rm erg~cm^{-2}~s^{-1}}$ for Chandra and $3\times10^{-15}~{\rm erg~cm^{-2}~s^{-1}}$ for XMM-Newton.

We have $M_{\rm BH}$ estimates from spectra acquired by the FMOS-COSMOS and zCOSMOS surveys with some AGNs having measurements from both. 
Considering the quality of the spectroscopy, the error on $M_{\rm BH}$ estimation, and the fact that ${\rm H}\beta$ line is used for calibrating virial mass estimators, we use the $M_{\rm BH}$ from ${\rm H}\beta$ (FMOS-COSMOS), ${\rm H}\alpha$ (FMOS-COSMOS), and Mg{\sc ii} (zCOSMOS-Bright, Deep) in order of preference; e.g., FMOS~${\rm H}\beta$ is used for an object with both FMOS~${\rm H}\beta$ and zCOSMOS~Mg{\sc ii} estimation. 
As shown in Figure~10 of \citet{Schulze2018}, there is a very good agreement between the FMOS H$\alpha$- and H$\beta$-based $M_{\rm BH}$ compared to those using MgII.
\cite{Schulze2018} also compared FWHM measurements taken at different times for each object (see Fig.7 of \citealt{Schulze2018}) and confirmed that the FWHM values are consistent with each other.
Note that 19 AGNs in our sample are listed in the SDSS DR16 quasar catalog \citep{Lyke2020}; we do not use these SDSS Mg{\sc ii}-based $M_{\rm BH}$ measurements given the benefits of the FMOS and deeper zCOSMOS spectroscopy. The number and redshift range of each measurement are summarized in Table~\ref{tab:sample}. 

\begin{table}[]
\caption{
Sample size for each single-epoch $M_{\rm BH}$ estimation
}\label{tab:sample}
\begin{tabular}{lllll}
\hline\hline
Survey & line & CW & PR & $z$ range\\ \hline
\multirow{2}{*}{FMOS-COSMOS} & ${\rm H}\alpha$ & 52 & 3 & 0.68--1.7\\
& ${\rm H}\beta$ & 21 & 2 & 1.2--2.5\\
zCOSMOS-bright & Mg~{\sc ii} & 26 & 0 & 1.1--2.1\\
zCOSMOS-deep & Mg~{\sc ii} & 3 & 0 & 0.94--1.2\\
Total & & 102 & 5 & 0.68--2.5\\
\hline
\end{tabular}
\end{table}


\begin{figure*}[ht!]\epsscale{1.15}
  \plotone{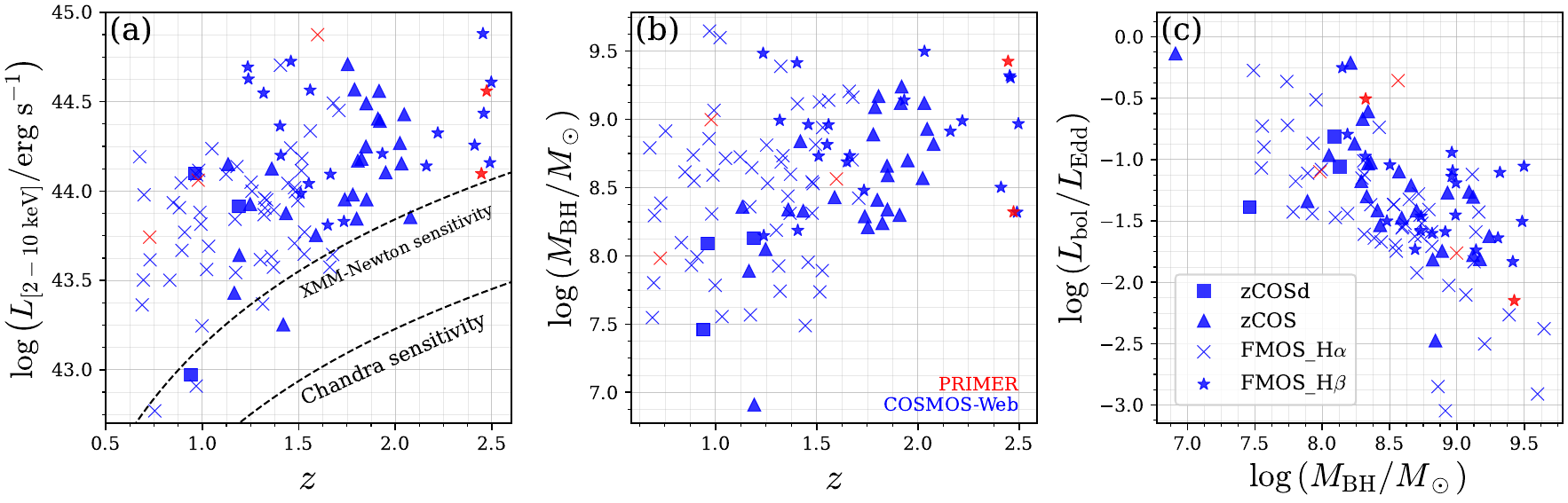}
  \caption{Characteristics of the type-I AGN sample with JWST imaging from COSMOS-Web and PRIMER:
  (a) Distribution of 2–10 keV X-ray luminosity $L_{\rm [2-10~keV]}$ as a function of $z$. Black dashed lines indicate the $F_{\rm [2-10~keV]}$ sensitivity in the sample, (b) Black hole mass $M_{\rm BH}$ as a function of $z$, and (c) relation between the Eddington ratio $L_{\rm bol}/L_{\rm edd}$ and $M_{\rm BH}$.
  The color and shape of the points indicate the JWST survey field and the source of $M_{\rm BH}$ estimation, respectively, as shown.
  \label{fig:sample_range}}
\end{figure*}

For ${\rm H}\beta$, \cite{Schulze2018} used the virial mass estimation relation by \cite{Vestergaard2006},
\begin{equation}
\small
    M_{\rm BH}\left({\rm H}\beta\right) = 10^{6.91} \left(\frac{L_{5100}}{10^{44}~{\rm erg~s^{-1}}}\right)^{0.5} \left(\frac{{\rm FWHM_{\rm H\beta}}}{1000~{\rm km~s^{-1}}}\right)^2 M_\odot, \label{eq:Hb}
\end{equation}
where $L_{5100}$ is continuum luminosity at 5100~\AA\ and ${\rm FWHM_{\rm H\beta}}$ is the full-width at half-maximum (FWHM) of ${\rm H\beta}$ broad line.
Then, the ${\rm H}\alpha$-based masses are calculated as given in \cite{Schulze2017, Schulze2018},
\begin{equation}
\small
    M_{\rm BH}\left({\rm H}\alpha\right) = 10^{6.71} \left(\frac{L_{{\rm H}\alpha}}{10^{42}~{\rm erg~s^{-1}}}\right)^{0.48} \left(\frac{{\rm FWHM_{\rm H\alpha}}}{1000~{\rm km~s^{-1}}}\right)^{2.12} M_\odot, \label{eq:Ha}
\end{equation}
$L_{{\rm H}\alpha}$  is the ${\rm H\alpha}$ luminosity and ${\rm FWHM_{\rm H\alpha}}$  is the FWHM of the broad ${\rm H\alpha}$ emission line.
For Mg{\sc ii}-based $M_{\rm BH}$ estimation (zCOSMOS-bright~Mg{\sc ii}, and zCOSMOS-deep~Mg{\sc ii}), the calibration by \cite{Shen2011} is used,
\begin{equation}
\small
    M_{\rm BH}\left({\rm Mg{\sc II}}\right) = 10^{6.74} \left(\frac{L_{3000}}{10^{44}~{\rm erg~s^{-1}}}\right)^{0.62} \left(\frac{{\rm FWHM_{Mg{\sc II}}}}{1000~{\rm km~s^{-1}}}\right)^{2} M_\odot, \label{eq:Mg}
\end{equation}
where $L_{3000}$ is continuum luminosity at rest-frame 3000~\AA\ and ${\rm FWHM_{Mg{\sc II}}}$ is FWHM of Mg{\sc ii} broad emission line.
These single-epoch virial mass estimations have an uncertainty due to possible variability and uncertainties in the modeling of broad-line regions \citep[c.f.,][]{Shen2013}.
In this paper, we use the $M_{\rm BH}$ uncertainties that also consider uncertainties from the single-epoch virial mass estimation, typically $\sim 0.4~{\rm dex}$.
We also consider this uncertainty in generating mock data (Section~\ref{subsubsec:_mock}).

From the parent catalog, we select broad-line AGNs that reside in the COSMOS-Web and PRIMER fields.
The final sample size has 107 AGNs with black hole mass estimation summarized in Table~\ref{tab:sample}.
Figure~\ref{fig:footprints} shows the spatial location of the AGN sample within the COSMOS-Web footprint.
We use five broad-band images from HST/ACS (F814W) and JWST/NIRCam (F115W, F150W, F277W, and F444W) for the sample residing in COSMOS-Web.
Four more broad-band (F090W, F200W, F356W) and medium-band (F410M) images are available for five galaxies in the PRIMER field.

We also compare the optical color index $g-i$ calculated based on the COSMOS2022 photometry \citep{Weaver2022} with SDSS quasars and hard-X-ray-detected AGNs at the same redshift range (Figure~6 in \citealt{Silverman2005}).
Our sample has a wide $g-i$ distribution of $g-i = 0 - 2.4$.
Thus, our sample includes both unobscured and dust-obscured AGNs and is not significantly biased to either sample.

Figure~\ref{fig:sample_range} shows the distribution of 2–10 keV X-ray luminosity $L_{\rm [2-10~keV]}$ (panel a) and $M_{\rm BH}$ (panel b) as a function of redshift, where $L_{\rm [2-10~keV]}$ is calculated with an X-ray spectral index $\Gamma=1.8$ \citep[e.g.,][]{Brightman2013}.
Since the sample consists of X-ray-selected objects and is flux-limited, there is a tendency for higher-$z$ objects to have larger $L_{\rm [2-10~keV]}$ over the sensitivity limit.
We can also see the trend of $M_{\rm BH}$ increasing with redshift.
The sample biases likely influence this trend from the flux sensitivity and the availability of broad-line FWHM measurements from spectroscopic data. Figure~\ref{fig:sample_range}~(c) displays the relation between the Eddington ratio $L_{\rm bol}/L_{\rm edd}$ and $M_{\rm BH}$ with Eddington ratio decreasing as $M_{\rm BH}$ increases.
This trend is due to the observational flux limitation and that $L_{\rm edd}$ is proportional to $M_{\rm BH}$.
While there is a correlation between $M_{\rm BH}$ and $L_{\rm bol}$, dividing $L_{\rm bol}$ by $L_{\rm edd}$ to calculate the Eddington ratio cancels out this weaker correlation between $L_{\rm bol}$ and $M_{\rm BH}$.

\begin{figure}[ht!]\epsscale{1.15}
  \plotone{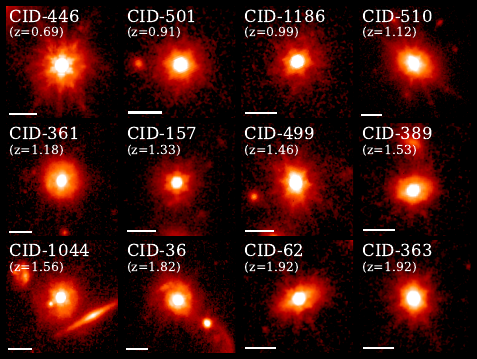}
  \caption{
  Original images in F277W of some targets in the order of redshift.
  The target IDs and the redshifts are shown in the left corner of each image.
  The white bars indicate a $1^{\prime\prime}$ scale.
  Depending on the host-to-total flux ratio ($H/T$) that varies with the sample, a central AGN component can dominate an entire system, thus burying a host galactic component.
  \label{fig:raw}}
\end{figure}

Figure~\ref{fig:raw} shows the original F277W images, i.e., before the decomposition analysis, of representative AGNs in our sample.
In some targets, we can recognize the extended components of the host galaxies far from the central PSF-like feature.
However, a central AGN component, especially those with spiky diffraction features in the outer part, dominates the system and buries the host galaxy component.
These dominant PSF components prevent us from obtaining host galaxy information directly and make the 2D decomposition analysis necessary.

\section{Method} \label{sec:method}
To extract host galaxy components from the original AGN + host galaxy NIRCam images, we apply a 2D image analysis tool {\tt galight} \citep{Ding2020_HST}.
With {\tt galight}, we perform forward modeling of each image as a superposition of a PSF component and PSF-convolved \sersic components corresponding to the light from an AGN and its host galaxy, respectively.
We then obtain images of the host galaxy, free of the AGN, by subtracting the fitted PSF component from the original image.

\subsection{Comparative analysis of model PSF construction}\label{subsec:psf}
Considering that the AGN can account for up to $\sim95\%$ of the total flux \citep[e.g.,][]{Ding2020_HST,Ding2022_CEERS}, the results significantly depend on the accuracy of reconstructing the PSF.
There are different strategies to reconstruct PSF images based on either using theoretical PSFs \citep[e.g.,][]{Suess2022} or stellar images \citep[e.g.,][]{Nardiello2022,Ding2022_CEERS,Ding2022_z6,Zhuang2023,Baker2023}.
The former uses the theoretical PSF model such as {\tt WebbPSF} \citep{Perrin2012,Perrin2014}, and the latter uses natural stellar images as the PSF directly or modeled PSF with tools such as {\tt PSFEx} \citep{Bertin2011}.
Many previous studies concluded that the synthetic PSF simulated by {\tt WebbPSF} is narrower than the PSFs reconstructed with the natural stars \citep{ono2022, Ding2022_CEERS, Onoue2023}.
Note that \cite{Ito2023} used the intermediate method between the former and the latter; they used theoretical PSFs with {\tt WebbPSF} and smoothed it by comparing the surface brightness profile of natural stellar images.
In this paper, we reconstruct the PSF using three methods based on natural star images for each region and filter.
Then, we compare the results with the different PSF reconstruction methods and discuss the host galaxy characteristics with the method dependence.

\subsubsection{$\chi^2_\nu$-based methods}\label{subsubsec:chisq_PSF}
First, we follow the strategy of \cite{Ding2020_HST,Ding2022_CEERS}.
They first construct an empirical PSF library for which each PSF is represented by the image of a single star.
Then, the 2D decomposition analysis is run with all single PSFs in the library, separately.
Then, they sort the results in the order of reduced chi-square $\chi^2_\nu$ and stack the PSFs with the top 3, 5, and 8 $\chi^2_\nu$ values.
Using the single PSFs and the stacked PSFs, they select the PSF with the smallest $\chi^2_\nu$ as the final PSF.
This method is based on the $\chi^2_\nu$; i.e., they assumed that the lower $\chi^2_\nu$ is indicative of a better (the closer to the more accurate) PSF.

Following their strategy, we apply the {\tt find\_PSF} function in {\tt galight} to list PSF candidates, then select PSF candidates manually for each mosaic image and filter.
In this manual selection process, only obvious PSF candidates with the PSF-like complex hexagonal and spiky diffraction features and without a galaxy-like broad component are selected.
We cropped the images of the selected PSF candidates for the short-wavelength-channel filters (F090W, F115W, F150W, and F200W) and long-wavelength-channel filters (F277W, F356W, F410M, and F444W) as squares with 150 and 240 pixels per side, corresponding to 4\farcs5 and 7\farcs2, respectively.
After removing neighbor objects using {\tt clean\_PSF} function in {\tt galight}, the PSF libraries contain $\sim30-50$ PSF candidates depending on the filter and region.
Then, we fit each AGN target with a superposition of a PSF-convolved \sersic profile and each single PSF candidate in the PSF library.
With the fitting results of each single PSF, we select PSFs with the top-5 $\chi^2_\nu$ and top-75\% $\chi^2_\nu$ and stack them to generate an averaged PSF image.
These top-5 stacked and top-75\% stacked PSFs are finally used to estimate the parameters in this method.
Note that each target has its own top-5 and top-75\% PSFs generated from single PSFs with the lowest $\chi_\nu^2$ selected for each target.

\subsubsection{Modeling method}\label{subsubsec:modeling_PSF}
\cite{Zhuang2023} compare JWST/NIRCam PSFs modeled with different methods ({\tt Swarm}, {\tt photutils}, and {\tt PSFEx}), and concluded that {\tt PSFEx} reconstructed PSFs provide the best performance.
From the 2D decomposition of simulated broad-line AGNs, \cite{Zhuang2023} also suggested that smaller $\chi^2_\nu$ values do not necessarily provide a means to distinguish which PSFs are more likely to characterize the AGN with higher accuracy.
Following the conclusion by \cite{Zhuang2023}, we use {\tt PSFEx} and compare the results with $\chi^2_\nu$-based selected PSFs (Section~\ref{subsubsec:chisq_PSF}).

{\tt PSFEx} constructs an empirical PSF model based on the output catalog of {\tt SExtractor} \citep{Bertin1996}.
We first run {\tt SExtractor} for source detection, and then run {\tt PSFEx} for modeling the PSF for each mosaic image and filter.
{\tt PSFEx} can also reconstruct local PSFs as a function of positions on the detector.
We do not use local PSFs because \cite{Zhuang2023} also concluded that a universal or global PSF usually shows ``satisfactory'' fitting results, and the sample region (COSMOS-Web and PRIMER-COSMOS) has a much smaller number of stars than the south continuous viewing zone, which \cite{Zhuang2023} tested local PSF reconstruction.

\begin{figure}[ht!]\epsscale{1.15}
  \plotone{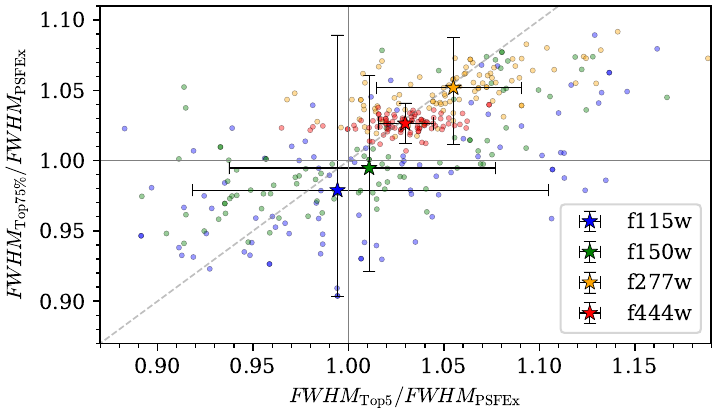}
  \caption{
  Comparison of FWHM (semi-major axis) for PSFs from different PSF reconstruction methods for all individual AGNs.
  The x-axis represents the ratio of the FWHMs for the top-5 PSFs to those for the {\tt PSFEx} PSFs, and the y-axis represents the ratio of the FWHMs for the top-75\% PSFs to those for the {\tt PSFEx} PSFs.
  Blue, green, orange, and red colors indicate the different filters: F115W, F150W, F277W, and F444W.
  Median values and $1\sigma$ confidence range are denoted by star symbols and error bars.
  The gray dashed line indicates $y=x$, i.e., the same FWHMs for the top-5 and top-75\% PSFs.
  Notably, F277W and F444W exhibit a FWHM bias among different PSF reconstruction methods.
  \label{fig:fwhm_comp}}
\end{figure}

\subsubsection{Comparing the final PSFs}\label{subsubsec:comp_finalPSFs}
Now, we have three final PSFs for comparison; the top-5 $\chi^2_\nu$ stacked PSF, the top-75\% $\chi^2_\nu$ stacked PSF, and the {\tt PSFEx} PSF (Sections~\ref{subsubsec:chisq_PSF} and \ref{subsubsec:modeling_PSF}).
To compare PSFs, we perform, a 2D Gaussian fitting for each PSF image and measure the FWHMs along the semi-major axis.
Figure~\ref{fig:fwhm_comp} compares the FWHMs of PSFs obtained with each method, target, and filter.
The x and y-axis of Figure~\ref{fig:fwhm_comp} show the ratio $FWHM_{\rm Top-5}/FWHM_{\rm PSFEx}$ and $FWHM_{\rm Top-75\%}/FWHM_{\rm PSFEx}$, where $FWHM$ is the value along the semi major axis.

Firstly, regardless of the filters, we can see that the distribution extends further in the x-axis direction than the y-axis.
This can be attributed to greater variation in the FWHMs for the top-5 stacked PSFs.
We use the same {\tt PSFEx} PSF for each target in the same field, and the top-75\% stacking in the same field uses mostly the same single PSFs in the field.
In contrast, top-5 stacking employs only the best-fit single PSFs with the lowest $\chi^2_\nu$.
As a result, the FWHM variation for each galaxy is largest for the top-5 PSF followed by the top-75\% PSF and {\tt PSFEx} PSF, and $FWHM_{\rm Top-5}/FWHM_{\rm PSFEx}$ have a larger scatter than $FWHM_{\rm Top-75\%}/FWHM_{\rm PSFEx}$.

Secondly, we focus on the FWHM bias between the methods for each filter.
As suggested by \cite{Zhuang2023}, for short-wavelength filters (F115W and F150W), there is a significant scatter in the FWHM ratio.
On the other hand, for long-wavelength filters (F277W and F444W), the scatter is smaller than the short-wavelength side.
These results imply that, in the long-wavelength filters,  {\tt PSFEx} PSFs are sharper than the top-5 and Top-75\% PSFs.
The impact of these trends on the 2D decomposition analysis is discussed in Section~\ref{subsub:diff_final_psf}.

Note that these trends can depend on the visual inspection performed when constructing the PSF library (Section~\ref{subsubsec:chisq_PSF}) and the settings used for Sextractor and {\tt PSFEx} (Section~\ref{subsubsec:modeling_PSF}).
For example, visual inspection can be biased by the hexagonal diffraction features of the JWST PSF, an appropriate FWHM range, and the absence of extended structures originating from host galaxies.
If this selection process is strongly biased by the hexagonal features, it might lead to a selective choice of brighter PSFs.
As a result, the parameter distributions presented here may not necessarily match the distribution of the actual PSF.
Nonetheless, even different PSF reconstruction methods can result in different FWHMs.
Therefore, when performing a 2D decomposition analysis with only one PSF reconstruction method and not considering the possibilities of other PSFs, 2D decomposition results can be biased by a specific PSF.
Because determining the PSF shape perfectly is challenging, it is also important to discuss uncertainties by considering the results obtained with possible different PSFs.
We discuss how different PSF reconstruction methods affect the results of the 2D decomposition and the final $M_*$ estimation in Section~\ref{subsub:diff_final_psf}, and we perform a detailed comparison of the obtained final PSFs in appendix~\ref{app:different_PSF}.

%

\begin{figure*}[ht!]\epsscale{1.15}
  \plotone{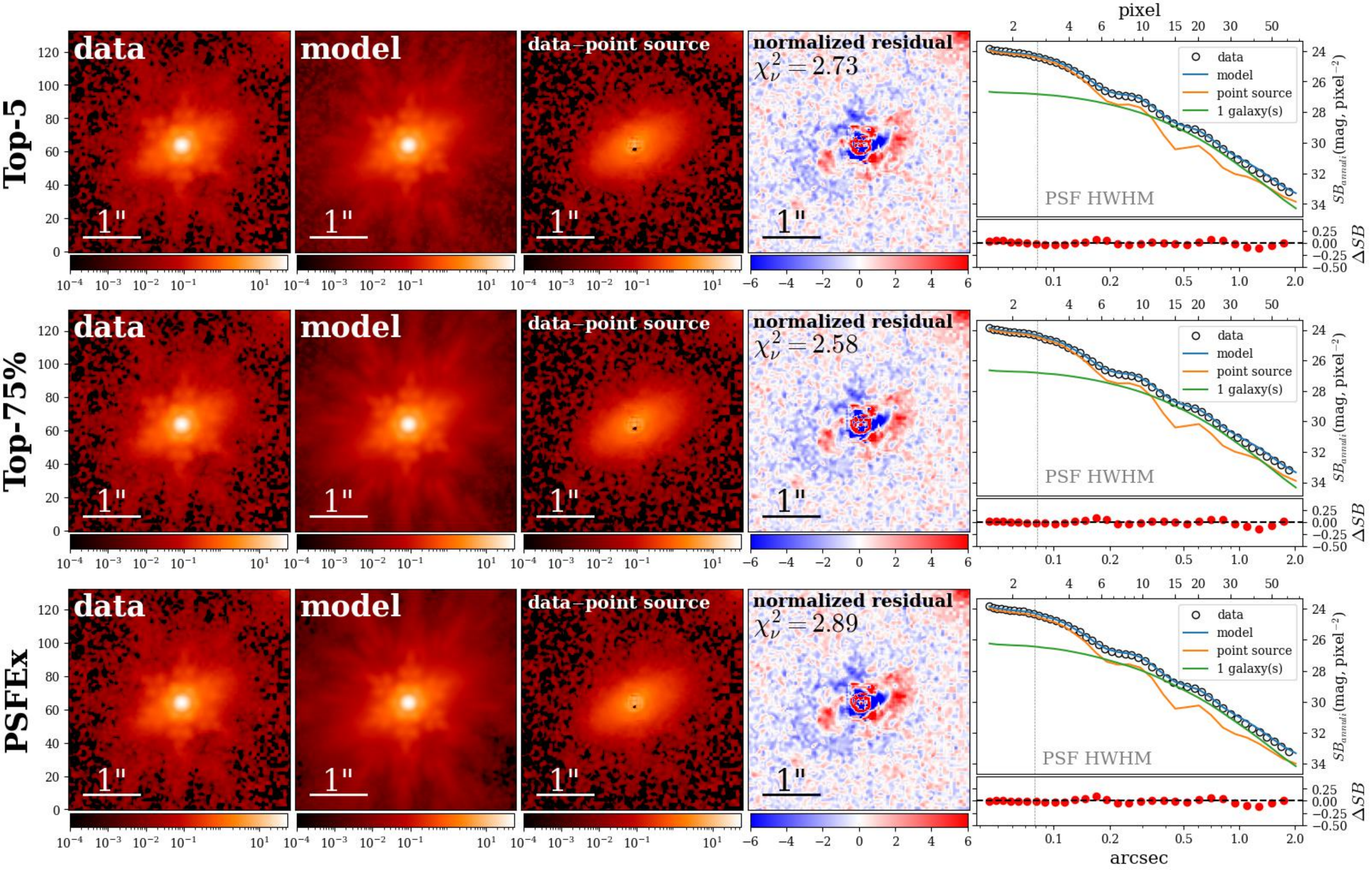}
  \caption{
  Example fits for CID-62 ($z\sim1.92$) using F444W.
  Each row shows the result for a different PSF (top-5 stacked, top-75\% stacked, and {\tt PSFEx}, from top to bottom).
  In each row, the images are as follows: original, model, data -- model point source (host galaxy only), and normalized residuals, from left to right. 
  $\chi_\nu^2$ values are shown in the panels of data -- model point source.
  The right panel shows the 1D surface brightness profile where dashed lines indicate half-width at half-maximum (HWHM) of each PSF.
  In the data-point source image, we reveal a disk-like host galaxy which is buried under the PSF before subtraction, regardless of the PSF reconstruction method.
  \label{fig:fitting_example}}
\end{figure*}

\subsection{Decomposition}\label{subsec:decomposition}
Using {\tt galight}, we fit the AGN + host galaxy images with the composite model of a PSF component and a single \sersic profile \citep{sersic1968} convolved by a PSF.
Note that we do not assume the same morphology in every band, i.e., the fitting is performed in each filter independently. 
In cases where there are nearby galaxies that can affect the fitting, these galaxies are also modeled as \sersic components and fitted simultaneously.
Our \sersic model has seven free parameters: amplitude, \sersic index $n$, effective radius $r_e$, coordinates of the center $x_c, y_c$, and ellipticity $e_1, e_2$.
The PSF model corresponding to the AGN component has three free parameters: amplitude and coordinates of the center $x_c, y_c$.
Thus, the number of free parameters for PSF + single \sersic component is ten in total.
Note that the actual number of free parameters in the fitting changes depending on the number of nearby objects also fitted with a \sersic profile.

To avoid unphysical results, the range of $n$ and $r_e$ is constrained to $\left[0.3, 7\right]$ and $\left[0\farcs06, 2\farcs0\right]$, respectively.
Note that some galaxies show clear substructures that do not suit a \sersic profile, such as bars and spiral arms.
Thus, using a \sersic profile is a first-order approximation to model the global component of AGN host galaxies.

In the fitting process, we cut the image into square regions centered on the target with a radius seven times the standard deviation along the semimajor axis of the 2D Gaussian fitting with {\tt photutils} \citep{photutils}.
Then, the above model is optimized with Particle Swarm Optimizer \citep[PSO;][]{kennedy1995}.
{\tt galight} also supports Markov Chain Monte Carlo (MCMC) to estimate the posterior parameter distributions.
As suggested by \cite{Ding2022_z6}, we also confirm that the uncertainty estimated with MCMC is much smaller than the uncertainties from different PSFs.
Thus, we do not use MCMC in the decomposition analysis, and we estimate the uncertainty from the results with different single PSFs (Section~\ref{subsec:photometry}).
We set the supersampling factor relative to pixel resolution to 3, which controls interpolation within a pixel to perform a subpixel shift of the PSF \citep[c.f.,][]{Ding2022_z6}.

As described above, using the PSF library constructed with {\tt galight}, we fit with every single PSF in the library and sort them in order of $\chi^2_\nu$.
Then we fit with the three final PSFs; top-5 $\chi^2_\nu$ stacked, top 75\% $\chi^2_\nu$ stacked, and {\tt PSFEx}.
Figure \ref{fig:fitting_example} compares the fitting results with the final PSFs for the example galaxy CID-62 in the F444W.
We can find that the host component is more prominent than the central PSF component at larger radii, and 2D decomposition makes it possible to detect the host galaxy initially buried under the PSF component with all three model PSFs.

For targets that have $n>6.5$ in any band other than F277W, we rerun the fit while fixing $n$ to the value found for the F277W band. 
F277W has the lowest number of values hitting the upper limit on $n$, falls above the rest-frame 4000~\AA\ break, and is somewhat central to JWST wavelength coverage.
This pertains to 42, 30, and 26 sources detected in F115W, F150W, and F444W, respectively.
From mock tests, we confirm that {\tt galight} can return $n\sim7$ even if the actual value is much smaller (Appendix~\ref{app:galight_mock}).
This is likely due to fitting where some of the AGN emission is attributed to a central stellar concentration of the host thus overestimating the host galaxy flux as well. 

\subsection{Detection of host galaxy}\label{subsubsec:detection}
Figure~\ref{fig:fitting_example} shows an example where the host galaxy is clearly detected.
However, for some galaxies, the strong PSF component dominates the total flux maybe due to not only a low host-to-total flux ratio $H/T$ but also compact morphology, making it challenging to distinguish the host signal from their PSF component.
To gauge, in a quantitative manner, which AGNs have accurate host galaxy information, we exclude cases where the host galaxy is undetectable, following three strategies below.

{\bf (1) \underline{Bayesian Information Criteria:}}
In addition to the PSF~+~\sersic model (PS+SE model) described above, we also fit with a model containing only a PSF component (PS model).
Then we calculate the Bayesian Information Criteria \citep[BIC;][]{Schwarz1978} for the two models, PS+SE and PS model, as,
\begin{equation}
    {\rm BIC} = \chi^2 + k\ln\left(n\right),
\end{equation}

\noindent where $k$ is the number of free parameters, and $n$ is the number of data points.
We regard that the PS+SE model provides a better description of the data than the PS model when ${\rm BIC_{PS+SE}}$ is much smaller than ${\rm BIC_{PS}}$, as
\begin{equation}
    {\rm BIC_{PS+SE}} < {\rm BIC_{PS}} - 10, \label{eq:con1}
\end{equation}
We decide the threshold value of the BIC difference of 10 based on \cite{Kass1995}.

{\bf (2) \underline{$S/N$ of the host galaxy:}}
To estimate the significance of the detection of the host galaxy, we calculate the $S/N$ of the host galaxy as done in \cite{Ding2022_z6}.
We construct an error map of the PSF-subtracted images considering two sources of error: noise from the observed images and the uncertainty propagated from different PSF reconstructions.
Due to the intrinsic variations of the PSF image even in the same FoV \citep{Zhuang2023, Yue2023} and errors in the observed image used in PSF reconstruction, the reconstructed PSFs contain uncertainties.
Thus, considering the uncertainty from PSF reconstruction is indispensable to calculate the $S/N$ of detected host galaxies.
For the PSF uncertainty, we use the pixel-by-pixel standard deviation of the fitted PSF components when fitting the PS+SE model with each single PSF in the PSF library.
Then, we calculate the final noise map as a composite of the observed noise map and the PSF uncertainty map.
With the final noise map, we calculate the signal-to-noise ratio of host galaxy $S/N_{\rm host}$ within a radius of 2$r_e$.
Then, we define the detection as cases with high $S/N_{\rm host}$, as 
\begin{equation}
    S/N_{\rm host} > 5. \label{eq:con2}
\end{equation}

{\bf (3) \underline{Manual inspection and removal:}}
We find that some objects have invalid central values in their F444W image and shallower surface brightness profiles than any PSFs.
The fitting of these galaxies fails even with applying a mask in the central region.
We find four cases (CID-50, CID-208, CID-668, and CID-112) with such features and label them as non-detections.

CID-142 is located near the edge of the image, and the host galaxy is partially cut off.
It is also the possible that CID-142 has a mismathced PSF from PSFs in other fields. 
Therefore, obtaining accurate photometry of CID-142 is challenging, thus we manually exclude CID-142 in the following discussion.

We also find some obvious false detections in F814W, where the image shows a dominant PSF feature and no extended host-like feature, and {\tt galight} fit the PSF-like features as a host galaxy with $H/T\sim1$. We confirm that the decomposition of the JWST images clearly detects a host-like extended feature.
Thus, in addition to the above strategies, we recognize 31 obvious false detections in F814W as non-detection cases.

With the above three strategies, we confidently report the detection of an AGN host galaxy for those that fulfill both the conditions~(\ref{eq:con1}) and (\ref{eq:con2}) for each of the final PSFs; i.e., we determine whether the host is detected or not for each top-5, top-75\%, and {\tt PSFEx} PSF, separately. 
Also note that this decision is made for each sample and each filter, i.e., a galaxy detected in one filter may not be detected in another filter.
The number of objects detected over two filters of NIRCam is
102, occupying $\sim 95\%$ out of the entire $N=107$ sample with the top-5 PSF (see Appendix~\ref{app:detnum} for the number of detection in each filter).

\subsection{Photometry of host galaxies}\label{subsec:photometry}
We calculate the flux of the host galaxies, using the \sersic fits, considering Galactic dust extinction \citep{Schlegel1998} for the detected cases.
For the photometric accuracy, we set an error of 0.2~mag, which represents likely systematic uncertainties \citep[e.g., ][]{Ding2022_CEERS, Zhuang2023, Zhuang2023COSMOS} and errors discussed in Section~\ref{subsubsec:detection}, considering both observational errors and PSF uncertainty in a radius of 2$r_e$.
We use the $3\sigma$ value for the undetected filters as the upper limit.

Here, we also fit with a model containing only a \sersic component (SE model).
For targets with ${\rm BIC_{PS+SE}} > {\rm BIC_{SE}} - 10$, the SE model is better or flexible enough to describe the data than the PS+SE model, and we use the \sersic photometry calculated with the SE model instead of the PS+SE model.
Such cases are observed only in F444W and are very limited (two or three objects depending on the PSF).


\subsection{SED fitting}\label{subsec:sed_fitting}
We fit the photometry of the host galaxies with {\tt CIGALE} ({\tt v2022.1}, \citealt{Boquien2019, Yang2022}) SED fitting library.
For a stellar population, we use the single stellar population model by \cite{Bruzual2003} ({\tt bc03} module) and the Chabrier initial mass function \citep[IMF;][]{Chabrier2003} with the $M_*$ cutoff of $0.1M_\odot$ and $100M_\odot$ \citep{Bruzual2003}.
We assume a delayed-$\tau$ model for a star-formation history (SFH), where SFR at each look-back time $t$ is modeled as 
\begin{equation}
    SFR\left(t\right) \propto
    \begin{cases}
        \left(t-t_{\rm age}\right) \exp\left(-\frac{t-t_{\rm age}}{\tau}\right) & \left(t>t_{\rm age}\right),\\
        0 & \left(t<t_{\rm age}\right),
    \end{cases}
\end{equation}
where $t_{\rm age}$ and $\tau$ indicate the starting time of star-formation activity and the declining timescale of SFR.
We also consider a nebular emission with {\tt nebular} module and a dust attenuation with {\tt dustatt\_modified\_starburst} module that assumes the modified \cite{Calzetti2000} law.
We set $E\left(B-V\right)$, $M_*$, $t_{\rm age}$, and $\tau$ as free parameters; their grid values are decided basically following \cite{Zhuang2023COSMOS} {\tt CIGALE} run and summarized in Table~\ref{tab:sed}.
To avoid unphysical solutions, we set the upper limit of $t_{\rm age}$ to $0.95 t_H$, where $t_H$ indicates the cosmic age at each redshift.
Otherwise, stellar metallicity and the ionization parameter $U$ are fixed at Solar metallicity and $\log U=-2$.

\begin{table*}[]
\scriptsize
\caption{Major parameter settings in SED fitting}\label{tab:sed}
\begin{tabular}{lp{4cm}p{4cm}p{4cm}}
\hline\hline
parameter & description & values ({\tt CIGALE}) & prior ({\tt Prospector}) \\ \hline
$\log\left(M_*/M_\odot\right)$ & total stellar mass & Scaled with the data & Uniform: min=9, max=13\\
$\log\left(Z_*/Z_\odot\right)$ & stellar metallicity & Fixed at 0 & Fixed at 0 \\
$\log U$ & ionization parameter & Fixed at -2 & Fixed at -2\\
$E\left(B-V\right)/{\rm mag}$ & Color excess for the nebular lines & 0, 0.001, 0.005, 0.01, 0.03, 0.05, 0.1, 0.2, 0.3, 0.4, 0.5, 0.6, 0.7 &  \\
$\hat{\tau}_{\rm dust,2}$ & optical depth of the diffuse dust attenuation &  & Uniform: min=0, max=3 \\
$\tau/{\rm Gyr}$ & timescale of delayed-$\tau$ SFH & 0.01, 0.05, 0.1, 0.25, 0.5, 0.75, 1.0, 1.25, 1.5, 2.0, 2.5, 3.0, 3.5, 4.0, 4.5, 5.0, 6.0, 7.0, 8.0, 9.0, 10, 12, 14, 16, 18, 20 & Uniform in $\log\left(\tau/{\rm Gyr}\right)$: min=-2, max=1.5\\
$t_{\rm age}$/Gyr & starting time of delayed-$\tau$ SFH & equally sampled in [1~Gyr, 1.95$t_H$] with the separation of 0.1~Gyr & Uniform: min=1, max=0.95$t_H$\\
\hline
\end{tabular}
\end{table*}

\begin{figure*}[ht!]\epsscale{1.0}
  \plotone{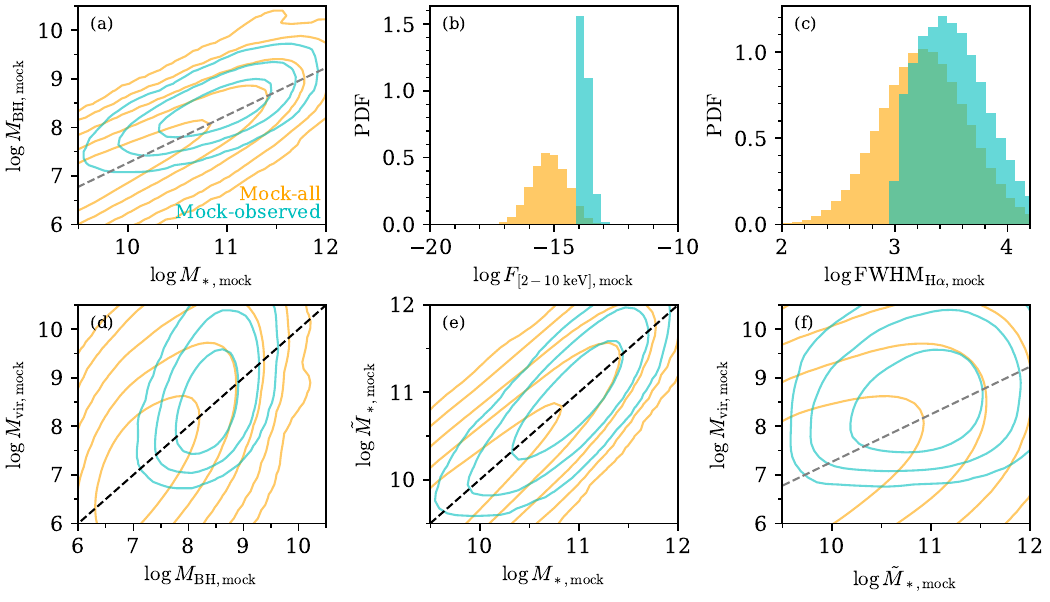}
  \caption{Procedure for generating mock galaxies at $z\sim1.5$ with an example parameter set of $\alpha=\alpha_{\rm local}$, $\beta=\beta_{\rm local}$, $\gamma=0$, and $\sigma_\mu=0.3$.
  Orange and blue contours show the distribution of the entire sample ($1\sigma-5\sigma$) and that above the detection limit of the observed sample ($1\sigma-3\sigma$, see Sections~\ref{subsec:sel_alpha} for the definition of the detection). The individual panels are as follows:
  (a) true black hole mass $M_{\rm BH, mock}$ vs. true stellar mass $M_{\rm BH, mock}$, before adding observational uncertainties,
  (b) $2-10~{\rm keV}$ flux ($F_{[2-10~{\rm keV}]), {\rm mock}}$),
  (c) ${\rm H\alpha}$ FWHM (${\rm FWHM}_{\rm H\alpha, mock}$),
  (d) pseudo-observed virial mass $M_{\rm vir, mock}$ vs. true black hole mass $M_{\rm BH, mock}$ where $M_{\rm vir, mock}$ is the mock black hole mass after considering the observational bias.
  (e) pseudo-observed stellar mass $\Tilde{M}_{\rm *, mock}$ vs. stellar mass $M_{\rm *, mock}$ where $\Tilde{M}_{\rm *, mock}$ is the stellar mass after considering the observational bias, and
  (f) pseudo-observed virial mass $M_{\rm vir, mock}$ vs. pseudo-observed stellar mass $\Tilde{M}_{\rm *, mock}$. The latter is to be compared with the observed $M_{\rm BH}-M_*$ plane shown in the left panel of Figure~\ref{fig:massmass}.
  Black dashed lines in panels (a) and (f) indicate $M_{\rm BH}-M_*$ local relation obtained from the fitting of the local galaxies \citep[][, see Section~\ref{subsec:mass-mass}]{Haring2004, Bennert2011}, and gray dashed lines in panels (d) and (e) indicate $y=x$ line.
  \label{fig:mockmake}}
\end{figure*}

We also apply the Bayesian-based spectral energy density (SED) fitting code, {\tt Prospector} \citep{Leja2017, Johnson2021}, to assess the uncertainty $M_*$ with different SED fitting codes.
{\tt Prospector} is based on the Flexible Stellar Population Synthesis ({\tt FSPS}, \cite{Conroy2009, Conroy2010}) to generate model SEDs of galaxies.
For comparison, we use almost the same settings with {\tt CIGALE}: a Chabrier IMF, a delayed-$\tau$ SFH model, Solar metallicity, \cite{Calzetti2000} dust attenuation law, and nebular emission with $\log U=-2$.
{\tt Prospector} can fit with the non-parametric SFH, which separates galaxy formation history into several age bins and assumes a constant SFR in each bin \citep[e.g.,][]{Leja2019}.
\cite{Lower2020} input cosmological hydrodynamic simulation data into {\tt Prospector} and concluded that non-parametric SFH tends to reconstruct $M_*$ more accurately than parametric SFHs.
\cite{Lower2020} also suggested that parametric SFH tends to underestimate $M_*$. 
However, in this study, the available photometry is limited in the number and wavelength range (only five/nine bands in the near-infrared wavelength range for the COSMOS-Web/PRIMER-COSMOS field).
Furthermore, the photometry derived in Section~\ref{subsec:photometry} contain uncertainties from the 2D decomposition analysis.
Therefore, we choose to use the parametric SFH (delayed-$\tau$ model) instead of the non-parametric assessment. 
The parameter prior settings in the MCMC run are summarized in Table~\ref{tab:sed}.
We compare the results with {\tt CIGALE} and {\tt Prospector} in Section~\ref{subsec:SEDfit_mass}.

\subsection{Generating mock data to consider selection effects}\label{subsubsec:_mock}

As mentioned, our sample is X-ray-flux limited (Section~\ref{subsec:sample_selection}), raising the possibility of bias toward larger $M_{\rm BH}$ or higher Eddington ratio \citep{Lauer2007,Schulze2011,Schulze2014}.
Due to this selection effect, a direct comparison of the observational results with the local relation is not appropriate.
Thus, in this study, we generate a mock AGN-galaxy catalog based on the procedure in \cite{Li2021_HSC} and apply the mock observation (adding selection biases and observational effects) to discuss the intrinsic evolution of the mass relation. The procedure for generating the mock catalogs is described below.

First, we generate the mock redshift $z_{\rm mock}$ and mock true stellar mass $M_{*,{\rm mock}}$ based on the COSMOS2020 stellar mass function (SMF) by \cite{Weaver2023}.
Next, we use the $M_{*,{\rm mock}}$ to generate the mock true BH mass $M_{\rm BH, mock}$ for the mock sample.
Here, we assumed the local relation as
\begin{equation}
    \log\left(\frac{M_{\rm BH}}{M_\odot}\right) = \alpha\log\left(\frac{M_*}{M_\odot}\right) + \beta,
\end{equation}
where $\alpha$ and $\beta$ indicate the slope and the intercept of the local $M_{\rm BH}-M_*$ plane.
Then, assuming a normal distribution, we calculate $M_{\rm BH, mock}$  as,
\begin{equation}
    \log M_{\rm BH,mock} = \mathcal{N}\left(\alpha\log\left(\frac{M_*}{M_\odot}\right) + \beta + \gamma\log\left(1+z\right), \sigma_\mu^2\right). \label{eq:mock_mr}
\end{equation}
Here, the parameter $\gamma$ indicates the strength of redshift evolution of the mass relation, and $\sigma_\mu$ is the intrinsic scatter of the mass relation.
Figure~\ref{fig:mockmake} (a) shows $M_{\rm BH,mock}$ vs. $M_{*, {\rm mock}}$ distribution with the example parameter set of $\gamma=0$ and $\sigma_\mu=0.3$.

Then, based on $z_{\rm mock}$ and $M_{\rm BH,mock}$, we assign mock Eddington ratio $\lambda_{\rm Edd, mock}$ by sampling the Eddington ratio distribution function by \cite{Schulze2015}.
Using the $\lambda_{\rm Edd, mock}$ and $M_{\rm BH,mock}$, we calculate the bolometric luminosity $L_{\rm bol, mock}$.

From $L_{\rm bol, mock}$, we obtain $L_{[2-10~{\rm keV}], {\rm mock}}$ using the bolometric correction by \cite{Duras2020}.
With the calculated $L_{[2-10~{\rm keV}], {\rm mock}}$, mock X-ray flux $F_{[2-10~{\rm keV}], {\rm mock}}$ is determined with the assumption of $\Gamma=1.8$, the same assumption as in Section~\ref{subsec:sample_selection}.
Figure~\ref{fig:mockmake} (b) shows $F_{[2-10~{\rm keV}], {\rm mock}}$ distribution with the example parameter set of $\gamma=0$ and $\sigma_\mu=0.3$.

We also generate mock virial BH mass $M_{\rm vir, mock}$ with the $L_{\rm bol, mock}$, $M_{\rm BH,mock}$, and the assumed FWHM distribution.
We first calculate mock continuum and line luminosity $L_{5100, {\rm mock}}$, $L_{3000, {\rm mock}}$, and $L_{{\rm H\alpha, mock}}$.
For $L_{5100, {\rm mock}}$ and $L_{3000, {\rm mock}}$, we used the bolometric correction by \cite{Netzer2007} and \cite{Trakhtenbrot2012}, respectively.
For $L_{\rm H\alpha, mock}$, we use the $L_{5100, {\rm mock}}$ and scaling relation between $L_{{\rm H\alpha}}$ and $L_{5100, {\rm mock}}$ by \cite{Jun2015}.
These bolometric correction and scaling relations are the same as those in \cite{Schulze2018}.
Then, we assume that FWHM of the emission line follows the log-normal distribution with the scatter of 0.17 dex \citep{Shen2008}, and we generate the observed FWHM of ${\rm H\alpha}$, ${\rm H\beta}$, and ${\rm Mg{\sc II}}$ emission lines following Equations~(\ref{eq:Hb}), (\ref{eq:Ha}), and (\ref{eq:Mg}).
To consider the bias in single-epoch virial mass estimation, we add the bias with $\beta_{\rm bias} = 0.6$ in calculating FWHM.
$\beta_{\rm bias}$ represents the proportion by which $\Delta L$, the variation of luminosity from the mean luminosity $\bar{L}$, affects the variation in FWHM.
Thus, the FWHM is generated by a lognormal distribution with the scatter of 0.17~dex and the mean value corresponding to the luminosity of $\bar{L} + \beta\Delta L$ \citep[c.f.][]{Shen2013, Li2021_EoR}.
Finally, the mock virial BH mass $M_{\rm vir, mock}$ is calculated using the mock observed FWHM and the luminosity following Equations~(\ref{eq:Hb}), (\ref{eq:Ha}), and (\ref{eq:Mg}).
Figures~\ref{fig:mockmake} (c) and (d) show the $FWHM_{\rm H\alpha, mock}$ histogram and $M_{\rm vir, mock}-M_{\rm BH,mock}$ distribution with the example parameter set of $\gamma=0$ and $\sigma_\mu=0.3$.

Considering possible systematic uncertainties from 2D decomposition analysis, we added an error based on a normal distribution with 0.2~dex to the $M_{*,{\rm mock}}$ to consider the uncertainty of $M_*$ derived from observations, resulting in the mock observed stellar mass $\Tilde{M}_{*,{\rm mock}}$.
The resulted $\Tilde{M}_{*,{\rm mock}}-M_{*,{\rm mock}}$ distribution is shown in Figure~\ref{fig:mockmake} (e).

Finally, we get the mock data set of the redshift $z_{\rm mock}$, observed stellar mass $\Tilde{M}_{*,{\rm mock}}$, and virial BH mass $M_{\rm vir, mock}$.
Figure~\ref{fig:mockmake} (f) shows the final mock observed mass distribution ($M_{\rm vir, mock} - \Tilde{M}_{*,{\rm mock}}$) with the example parameter set of $\gamma=0$ and $\sigma_\mu=0.3$.
In Section~\ref{subsec:mass-mass}, we discuss the mass relation, comparing our results with this mock catalog.

\begin{figure*}[ht!]\epsscale{1.15}
  \plotone{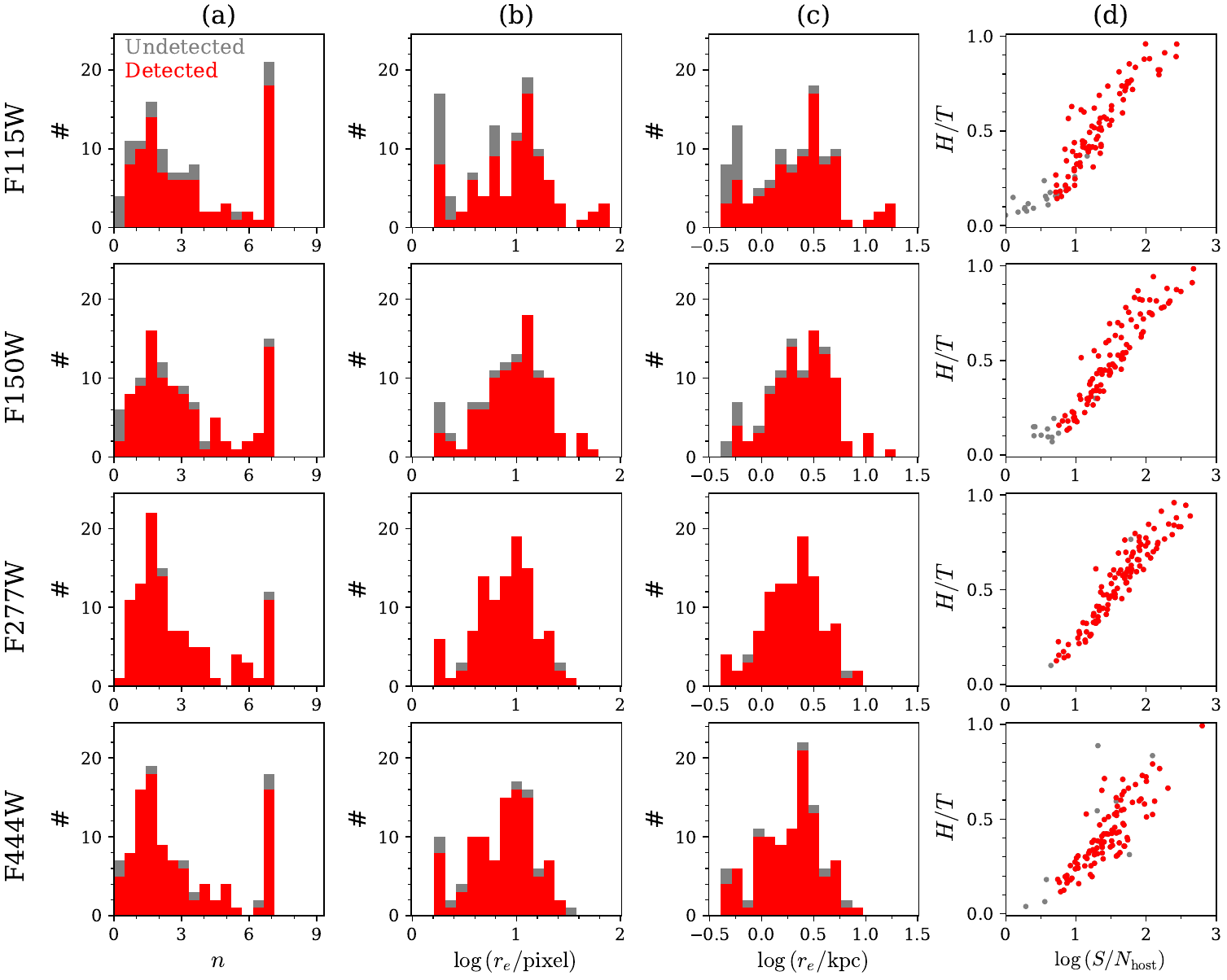}
  \caption{
  Summary of the decomposition results in each filter.
  Columns~(a), (b), and (c) respectively show the distribution of the estimated $n$, $\log r_e$ in the unit of pixel and kpc.
  In column~(d), the relation between $H/T$ and $S/N_{\rm host}$ is shown.
  In each panel, red (gray) indicates detection (non-detection) in each filter.
  \label{fig:param_dist}}
\end{figure*}

\section{Results of AGN - host decomposition}\label{subsec:morphology}

\subsection{Morphological parameters}
Firstly, we examine the results obtained from the 2D decomposition. 
In Figure~\ref{fig:param_dist} (a)--(c), we show the distribution of $n$ and $r_e$ (in the unit of pixel and kpc) for each filter using the top-5 stacked PSF. The distribution of $n$ has a peak in the distribution at $n\sim1-2$, clearly seen in the F277W and F444W filters. 
This is similar to studies of AGN hosts at high redshift, which show hosts characterized by disk-like morphology. 

In Figure~\ref{fig:param_dist} (d), we compare $H/T$ with $S/N_{\rm host}$.
We can see a strong correlation between the reconstructed $H/T$ and $S/N_{\rm host}$.
High $H/T$ means host galaxies dominate the AGN + host galaxy composite images, and we can easily detect host galaxies with high $S/N_{\rm host}$; thus, our results indicate the validity of our analyses for the majority of the sample.
Estimated morphological parameters and basic information for each host galaxy are reported in Table~\ref{tab:data_summary}.
In Appendix~\ref{app:galight_mock}, we confirm that {\tt galight} can reconstruct $r_e$ and $H/T$ correctly by running {\tt galight} on mock galaxy images.

\begin{figure*}[ht!]\epsscale{1.15}
  \plotone{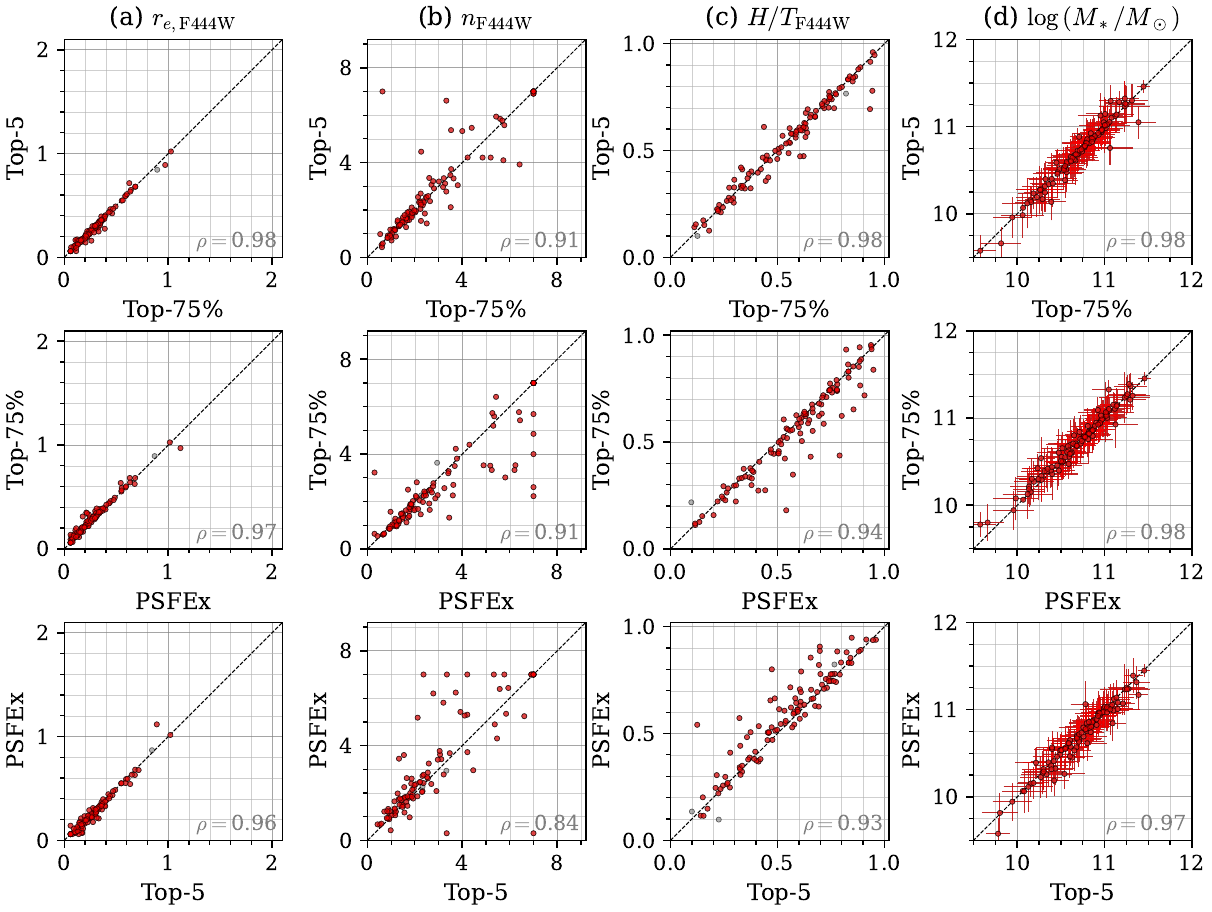}
  \caption{Comparison of the decomposition results ($r_e$ in the unit of arcsec, $n$, and $H/T$) based on the F444W filter using different PSFs (Columns~a--c). Data points in red have satisfied our stringent criteria for host detection (Sec.~\ref{subsubsec:detection}).
  Column~(d) compares $M_*$ estimated by the SED fitting, with colors corresponding to whether each object was detected in more than two bands.
  Spearman's correlation coefficient $\rho$ for each prior is shown in the lower right corner of each panel.
  High correlation coefficients and the distribution around $y=x$ (black dashed line) suggest consistent results among the fitting with different PSFs.
  \label{fig:comp_res_deco}}
\end{figure*}

\begin{figure*}[ht!]\epsscale{1.1}
  \plotone{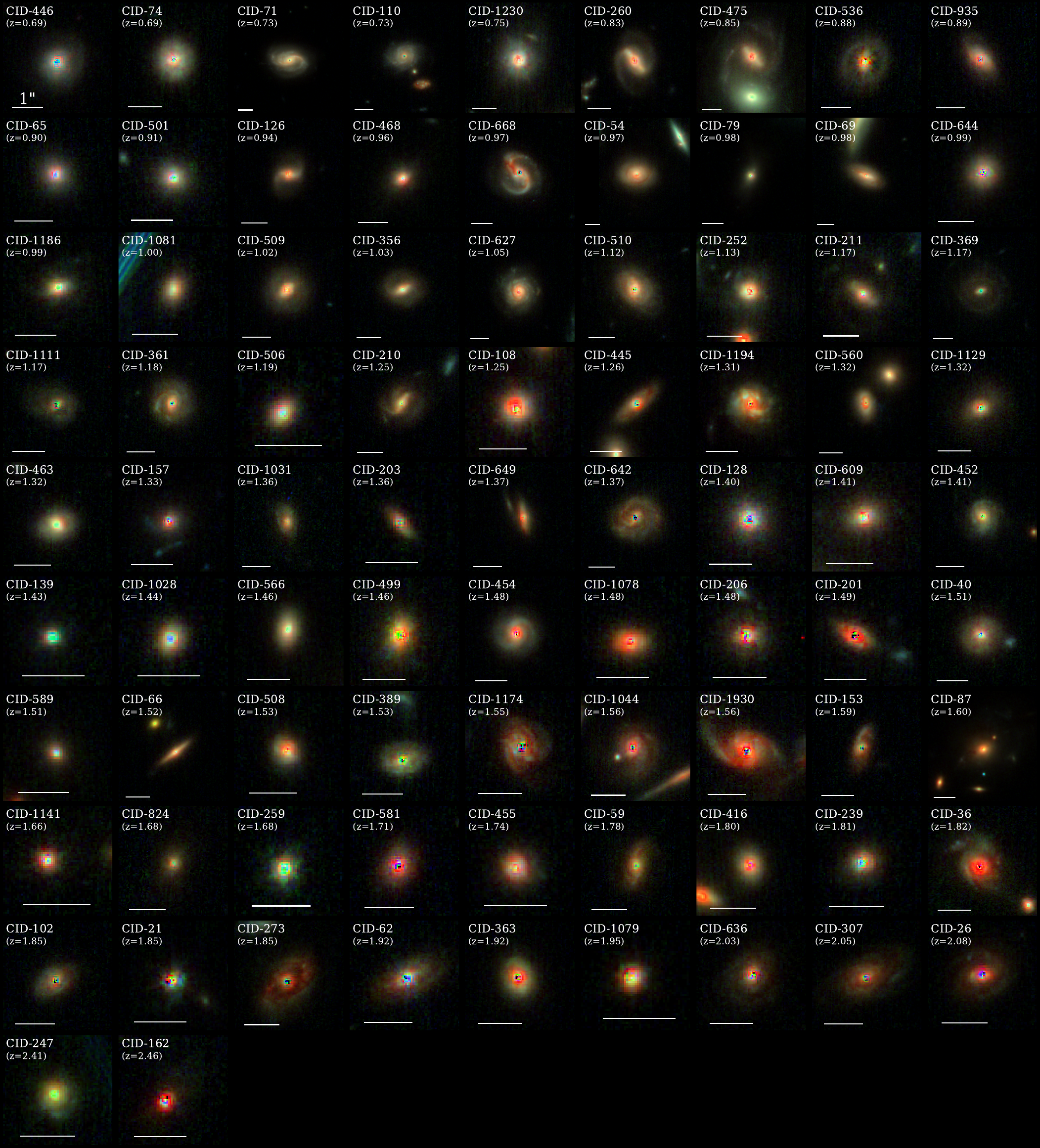}
  \caption{
  Three-color cutout PSF-subtracted images (the F277W, F150W, and F115W for RGB) of all host galaxies detected in all F115W, F150W, and F277W in the order of redshift.
  We have not performed deconvolutions based on the PSF FWHM differences between each filter, and consequently, each filter image possesses a different-sized PSF.
  The target IDs and the redshifts are shown in the up-left corner of each image.
  The white bars in the lower left part are $1^{\prime\prime}$ in length.
  Thanks to the high spatial resolved deep observation of JWST and the careful decomposition analysis, we can clearly identify substructures, such as bulges, spirals, bars, and dust lanes.
  \label{fig:3col}}
\end{figure*}

\subsection{Impact of different PSF models}\label{subsub:diff_final_psf}
In Figure~\ref{fig:comp_res_deco} (a)--(c), we compare morphological parameters, $n$, $r_e$, and the host-to-total flux ratio $H/T$, in F444W obtained using each final PSFs.
These parameters fall mostly along the $y=x$ line, indicating a strong correlation.
Thus, we can say that different PSF reconstruction methods do not significantly affect the results.

In Figure~\ref{fig:comp_res_deco} (a), the size ($r_e$) comparison shows more consistent results than $n$, indicating that it is minimally affected by different PSFs.
On the other hand, in comparing $n$ (Figure~\ref{fig:comp_res_deco} (b)), regardless of the PSF, we can find a consistent estimation on the low $n$ side ($n\lesssim3$).
However, at larger $n$ ($\gtrsim4$), the scatter increases. 
This tendency is likely because larger $n$ implies a compact \sersic profile that resembles the PSF, making it challenging to distinguish from the PSF.
We also find a tendency that {\tt PSFEx} PSFs result in slightly larger $n$ than the top-5 or top-75\% stacked PSFs. 
Even when considering the other filters, we find highly consistent results, with a trend of increased scatter at higher $n$.

Regarding $H/T$, we primarily see consistent results with strong correlations in Figure~\ref{fig:comp_res_deco} (c).
While $H/T$ estimated using the top-5 and top-75\% PSFs is largely consistent with each other, the {\tt PSFEx} PSFs tend to result in slightly higher $H/T$ than the other two PSFs.
\cite{Zhuang2023} suggested that using a narrower PSF than an exact PSF could overestimate the host flux and $n$.
Thus, the above biases in $n$ and $H/T$ could be explained by the fact that {\tt PSFEx} PSFs for F444W have a slightly narrower PSF than the other PSFs, as shown in Figure~\ref{fig:fwhm_comp}.
A detailed comparison of the estimated $H/T$ in other filters is summarized in Appendix~\ref{app:different_PSF}.

In conclusion, the decomposition results using different PSFs are generally highly consistent with each other.
A comprehensive discussion of the technical and practical differences between PSF reconstruction methods will also be provided in Section~\ref{subsec:PSF_dependence}.
Nonetheless, these discussions are based on the comparisons between estimated values, and here, we cannot definitively determine the true exact value.
Related to this, we provide the result of mock tests using different final PSFs in Appendix~\ref{app:galight_mock}.
Also, note that all final PSFs are not single stellar images but stacked or modeled PSFs based on multiple stellar images. 
As mentioned in Section~\ref{subsubsec:comp_finalPSFs}, single PSFs often have a wide range of sizes and shapes.
Thus, using a single stellar image without testing other stellar images can risk misinterpreting the PSF image and giving different results.

\subsection{Host images}\label{subsec:hostimage}
As seen in these high-quality host galaxy images, 2D decomposition analyses of JWST images open up the potential for a more detailed image-based galaxy analysis, such as studies traditionally conducted on inactive galaxies.
Figure~\ref{fig:3col} shows three-color images (F277W, F150W, and F115W for RGB) of each host galaxy created by subtracting the PSF and the nearby Sersic component.

Firstly, thanks to the high spatial resolution, deep observations, and meticulous decomposition analysis, we can access the highest-quality AGN-host galaxy images up to $z\sim2.5$, allowing us to identify substructures.
Particularly, in the case of CID-273 ($z=1.85$) and CID-307 ($z=2.05$), despite their redshifts $\sim2$, we can clearly identify blue spiral arms with an overall diffuse red broad component.

Additionally, galaxies such as CID-54 ($z=0.97$), CID-510 ($z=1.12$), CID-361 ($z=1.18$), CID-445 ($z=1.26$), and CID-452 ($z=1.41$) show more reddish colors at their centers than in the outer region, indicating the possibility of having a bulge-like structure or highly dust-obscured region (e.g., \citealt{Ito2024}).
Furthermore, there are cases with extended red structures (e.g., CID-668; $z=0.97$) which may indicate the presence of dust lanes as seen in the X-ray obscured (type-II) AGNs \citep{Silverman2023}.


\subsection{Estimated $M_*$ and the comparison of different SED fitting methods}\label{subsec:SEDfit_mass}
\begin{figure*}[ht!]\epsscale{0.9}
  \plotone{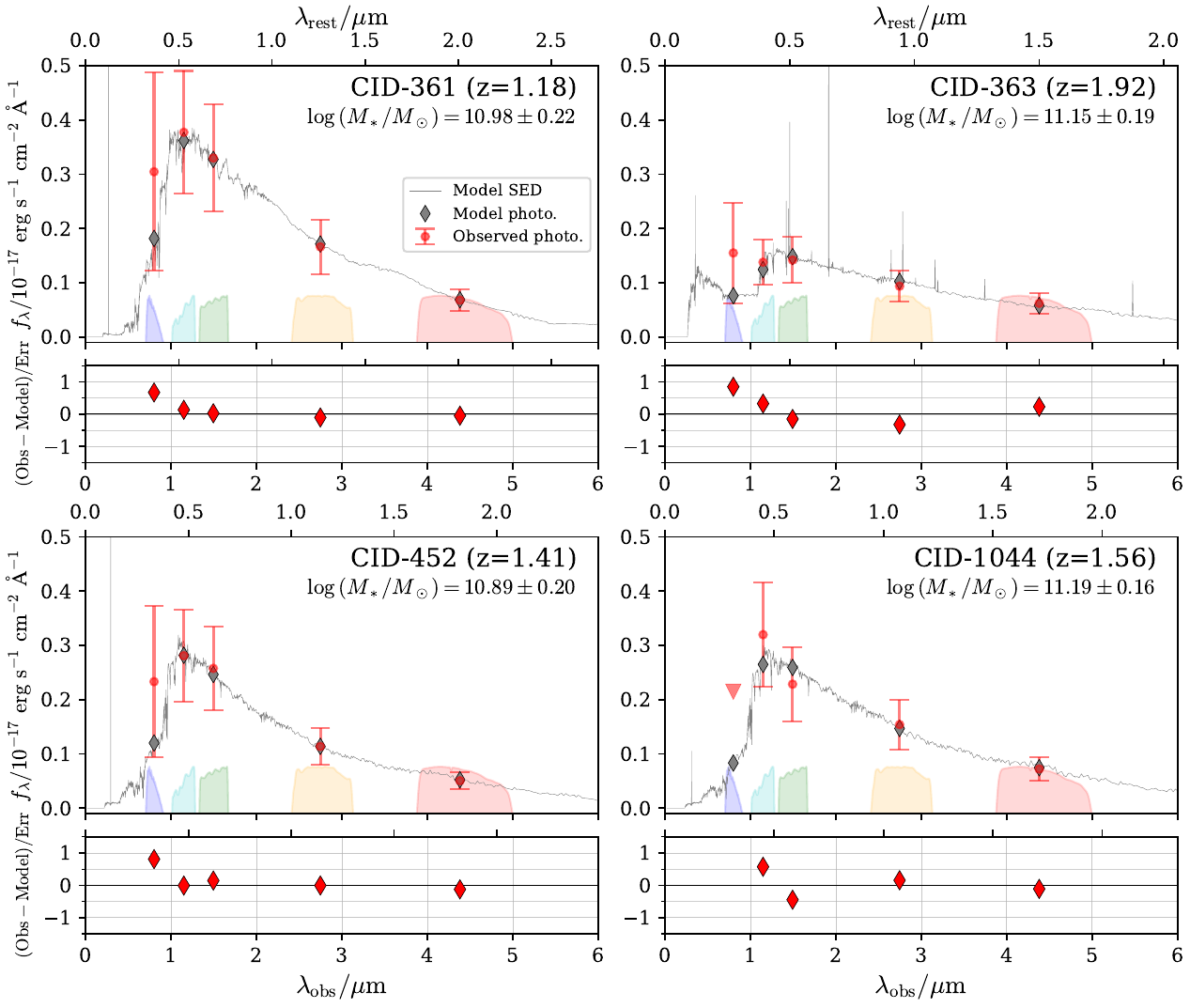}
  \caption{Representative examples of SED fits to PSF-subtracted host galaxies. 
  Source ID and their estimated $M_*$ with 1$\sigma$ confidence range are shown in the top right corner of each plot.
  The inverted triangle indicates the $3\sigma$ upper limit due to the non-detection in each filter.
  The bottom colored shades show the transmission curves of F814W, F115W, F150W, F277W, and F444W from left to right.
  The lower plot in each panel shows the difference between the observational and best-fit model photometry scaled by the error of the data.
  \label{fig:spec}}
\end{figure*}

As described in Section~\ref{subsec:sed_fitting}, we perform SED fitting of the host galaxy photometry to estimate $M_*$.
For five galaxies detected in less than two bands, we estimate $M_*$ upper limits by assuming $H/T=0.2$ and the average F277W photometry-to-$M_*$ ratio in the sample.
The assumed $H/T$ of $0.2$ is because galaxies with $H/T\gtrsim0.2$ are almost detected in our methods (Figure~\ref{fig:param_dist}(d)).
The inferred $M_*$ for each host galaxy is reported in Table~\ref{tab:data_summary}.
Figure~\ref{fig:spec} shows four examples of the SED fitting results with the residuals.
Generally, for the objects with $z\gtrsim1$, the photometry or upper limit from F814W and F115W fall at a rest-frame wavelength shorter than 4000~\AA\ break and are important in constraining stellar population parameters of host galaxies.

\begin{table*}
    \caption{Data summary of our sample.\tablenotemark{$^*$}}\label{tab:data_summary}
    \begin{tabular}{lllllllllll}
    \hline\hline
        CID & XID & $z$ & R.A.\tablenotemark{$^a$}  & Decl.\tablenotemark{$^a$}  & $\log M_{\rm BH}$\tablenotemark{$^b$}  & line\tablenotemark{$^c$} & $\log M_*$\tablenotemark{$^d$} & $r_{e,{\rm F115W}}$\tablenotemark{$^e$} & $n_{\rm F115W}$\tablenotemark{$^e$} & $\log H/T_{\rm F115W}$\tablenotemark{$^e$} \\ 
        & & & [deg] & [deg] & $[M_\odot]$ & & $[M_\odot]$ & [$^{\prime\prime}$] & &\\ \hline
        157 & 5404 & 1.33 & 149.6751 & 1.9828 & 8.73 & 2 & $10.43\pm0.24$ & 0.27 & 5.2 & -0.38\\ 
        162 & 169 & 2.46 & 149.7359 & 2.0276 & 9.30 & 1 & $11.28\pm0.17$ & 0.06 & 7.0 & -0.57 \\ 
        416 & 404 & 1.80 & 149.8129 & 2.3455 & 8.41 & 3 & $10.98\pm0.13$ & 0.22 & 1.9 & -0.36\\ 
        203 & 341 & 1.36 & 149.8142 & 2.0164 & 8.34 & 3 & $10.18\pm0.25$ & 0.18 & 1.5 & -0.37\\ 
        307 & 285 & 2.05 & 149.8227 & 2.0897 & 8.93 & 3 & $11.08\pm0.21$ & 1.5 & 7.0 & -0.39\\ 
    \hline
    \end{tabular}
    \tablenotetext{$*$}{The full version of this catalog is to be available online which will include results based on all bands.}
    \tablenotetext{$a$}{J2000}
    \tablenotetext{$b$}{$M_{\rm BH}$ is estimated with single epoch virial mass estimation using each line. We assume the 0.4~dex uncertainty in this study.}
    \tablenotetext{$c$}{1:${\rm H\beta}$ (FMOS-COSMOS), 2:${\rm H\alpha}$ (FMOS-COSMOS), 3:Mg~{\sc ii} (zCOSMOS-bright), 4:Mg~{\sc ii} (zCOSMOS-deep)}
    \tablenotetext{$d$}{$M_*$ is estimated using {\tt CIGALE} with the top-5 PSF results.}
    \tablenotetext{$e$}{Table~\ref{tab:data_summary} only shows morphological parameters in F115W measured with the top-5 PSF.
    Parameters in the other filters and with the other final PSFs are to be available in the full version.}
\end{table*}

\begin{figure*}[ht!]\epsscale{1.15}
  \plotone{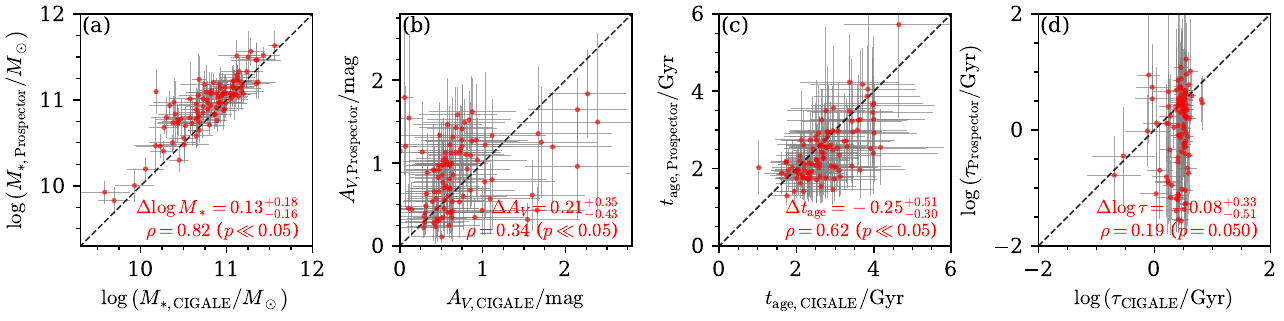}
  \caption{
  Comparison of SED fitting results with two different SED fitting codes. 
  Panels~(a) to (d) compare the estimated $\log M_*$, $A_V$, $t_{\rm age}$, and $\log\tau$.
  X and Y axis correspond to the result with {\tt CIGALE} and {\rm Prospector}.
  The black dashed line indicates the $x=y$ line.
  The median and $1\sigma$ confidence level of the difference between the two results (defined as $\Delta={\rm Prospector} - {\rm CIGALE}$) and Spearman's correlation coefficient are shown in the lower right part in each panel.
  \label{fig:pro_cig}}
\end{figure*}

We independently employ two distinct SED fitting codes, {\tt CIGALE} and {\tt Prospector}, as explained in Section~\ref{subsec:sed_fitting}.
Both codes are run having as similar parameter settings as possible. 
In Figure~\ref{fig:pro_cig}, we compare the parameters obtained from both codes. As shown in Figure~\ref{fig:pro_cig} (a), the results exhibit a significantly high positive correlation. 
However, we find an offset of approximately 
$\Delta \log M_* =+0.13~{\rm dex}$ (corresponding to $1\sigma$).
This offset is not far from a common systematic $M_*$ uncertainties among SED fitting methods reported in \cite{Pacifici2023}.
It remains challenging to determine whether $M_*$ from {\tt CIGALE} or {\tt Prospector} is more accurate.
In this study, we primarily use the {\tt CIGALE} $M_*$ in the main discussion to maintain consistency with previous studies for AGN-host galaxies \citep[e.g.,][]{Zou2019, Ishino2020, Shen2020, Li2021_HSC, Koutoulidis2022, Zhuang2023COSMOS, Li2024}.

Finally, in Figure~\ref{fig:comp_res_deco} (d), we compare the estimated $M_*$ with each final PSF. The estimated $M_*$ are on the 1:1 line and show a strong correlation, suggesting that the estimated $M_*$ with different final PSFs remain highly consistent.
This consistency can be attributed to the fact that, as shown in Figure~\ref{fig:comp_res_deco} (d), $H/T$ or host flux from different final PSFs is also consistent. 

We compare our {\tt CIGALE} $M_*$ with \cite{Zhuang2023COSMOS}, which also utilized {\tt CIGALE} on the COSMOS-Web data, and find that these two measurements well agree with each other with a scatter of $\Delta\log M_*=+0.08^{+0.20}_{-0.18}$.
This is very encouraging since their decomposition analysis is independent of our effort.

In Figures~\ref{fig:pro_cig} (b)--(d), we compare results for the other output SED model parameters: $A_V$, $t_{\rm age}$, and $\tau$. 
While $A_V$ and $t_{\rm age}$ exhibit large uncertainties, they show a positive correlation, with the median offset being close to zero within the range of uncertainties. 
Regarding $\tau$, it is evident that significantly inconsistent values are observed around $\log\left(\tau/{\rm Gyr}\right)\sim0.5$. 
Considering that $t_{\rm age}$ is generally $\log\left(t_{\rm age}/{\rm Gyr}\right)\lesssim0.6$ in our sample, this discrepancy can be attributed to the challenge of accurately determining SFH when $\tau\gg t_{\rm age}$. 

Additionally, we confirmed a strong negative correlation ($\rho=-0.60$, $p\ll0.05$) between $\Delta\log\tau$ and $\Delta\log M_*$.
This relation is because larger $\tau$ indicates a longer-lasting SFH, i.e., star formation has been persisting more recently.
Consequently, galaxies with larger $\tau$ tend to host more young stellar populations, resulting in a smaller mass-to-light ratio and smaller $M_*$.
While these differences may be attributed to differences in stellar models or fitting strategies (i.e., Bayesian or non-Bayesian), further investigation is omitted in this paper since it does not impact the main results of this study.

\section{$M_{\rm BH}-M_*$ relation}\label{subsec:mass-mass}

\begin{figure*}\epsscale{1.15}
  \plotone{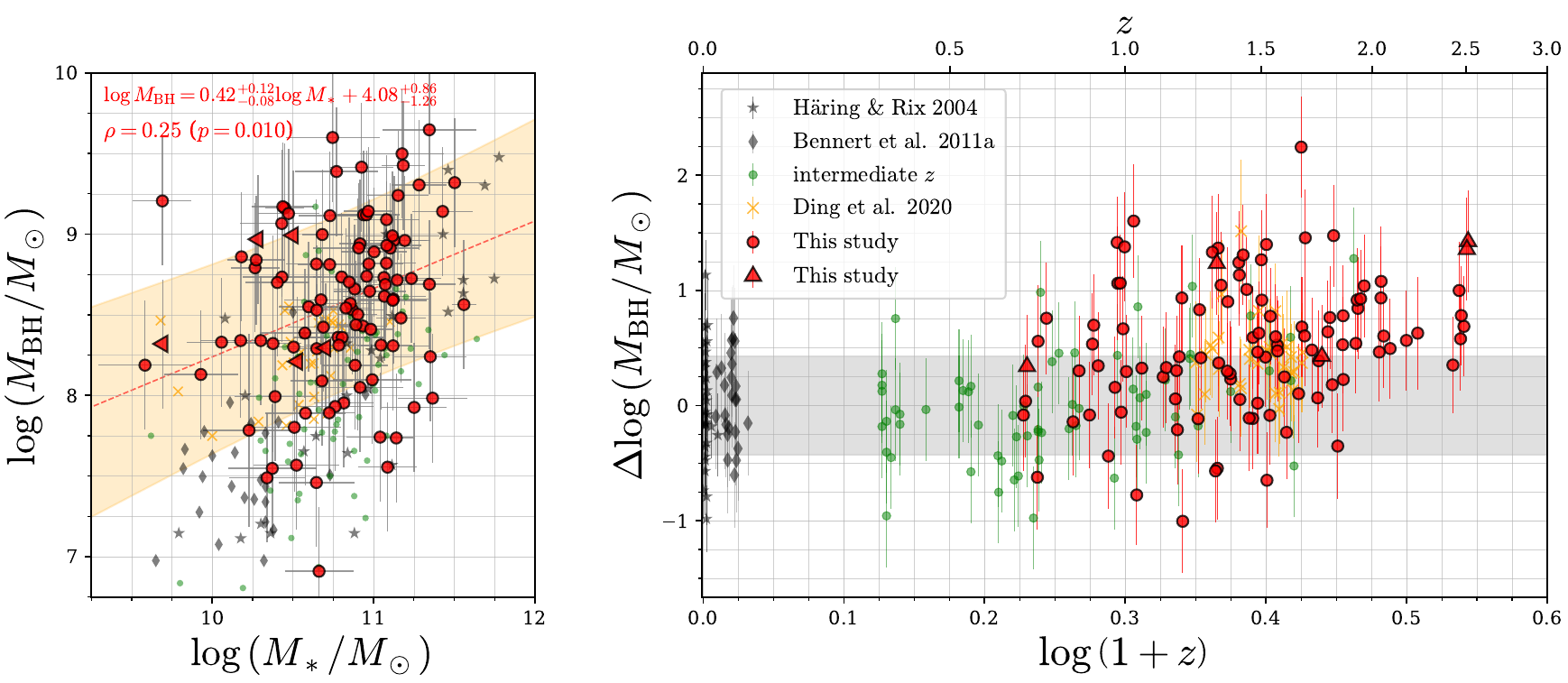}
  \caption{
  ({\it left}) Observed $M_{\rm BH}$ and $M_*$ distribution as shown by the red circles. A linear fit is indicated by the red line with the shaded region indicating the $1\sigma$ confidence range.  
  The best-fit relation and Spearman's correlation coefficient are reported at the top of the panel. 
  ({\it right}) $\Delta\log\left(M_{\rm BH}/M_\odot\right)$ as a function of redshift where zero corresponds to the local relation as marked by the gray band having a width representing the local dispersion.
  In both panels, samples from previous studies using HST are plotted: gray for low-$z$ \citep{Haring2004, Bennert2011}, green for intermediate-$z$ \citep{Jahnke2009,Bennert2011highz,Cisternas2011,Schramm2013} and orange indicating those from \cite{Ding2020_HST} at $z\sim1.5$.
  Red triangles indicate the $M_*$ upper limit and $\Delta\log\left(M_{\rm BH}/M_\odot\right)$ lower limit for five undetected targets.
  \label{fig:massmass}}
\end{figure*}

In the left panel of Figure~\ref{fig:massmass}, we plot our measurements of $M_{\rm BH}$ as a function of $M_{*}$.
Based on the large sample covering a broad range in both parameters, we find a weak positive correlation with a weighted Spearman's correlation coefficient\footnote{The weighted Spearman's 
 correlation function is calculated with R/CRAN {\tt wCorr} package \citep{wCorr}} of $\rho=0.25$.
As done in previous local studies, we attempt to fit the observational data with a linear function of 
\begin{equation}
    \log M_{\rm BH} = \alpha \log M_* + \beta. \label{eq:linear}
\end{equation}
The red line represents the results with $\alpha\simeq0.42^{+0.12}_{-0.08}$ and $\beta\simeq4.08^{+0.86
}_{-1.26}$, and the orange-shaded region showing the $1\sigma$ confidence interval.
The coefficient of determination $R^2$ for this linear fitting is 0.049, which indicates that this linear model does not sufficiently explain the data.
We test the validity of this linear model by fitting the data with a model with the fixed $\alpha$ of 0 and compared the BIC values.
Although the BIC for the $\alpha$-free model is lower than the BIC for the $\alpha=0$ model, the difference in BIC is less than 10.
This means the $\alpha$-free model is not significantly better than the $\alpha=0$ model.
We also run the Shapiro-Wilk test and got a p-value of 0.37.
Therefore, at a 5\% significance level, the null hypothesis cannot be rejected, and we cannot say that the observed data does not follow a Gaussian distribution.
However, it is important to note that this result does not consider the selection bias and does not necessarily represent the intrinsic distribution on the $M_{\rm BH}-M_*$ plane at high redshift.

For comparison, we also provide the linear fit to the local sample consisting of 30 inactive galaxies \citep{Haring2004} and 25 active galaxies \citep{Bennert2011}.
The fit to these 55 galaxies results in $\alpha_{\rm local}\simeq0.97^{+0.10}_{-0.11}$ and $\beta_{\rm local}\simeq-2.48^{+1.15}_{-1.11}$.
$R^2$ for this fitting is 0.63, and this suggests that the local $M_{\rm BH}-M_*$ relation is well described by the linear model.
Note that the local sample and our high-$z$ sample have different selection effects. Thus, we cannot directly compare $\alpha$ and $\beta$ with $\alpha_{\rm local}$ and $\beta_{\rm local}$ (see Sections~\ref{subsec:sel_alpha} and \ref{subsec:redshift_evolution} for the discussion with consideration of selection bias).

For investigating the redshift dependence of the mass relation, we calculate $\Delta\log\left(M_{\rm BH}/M_\odot\right)$, the relative offset of the black hole mass at given $M_*$ from the local relation:
\begin{equation}
    \Delta\log\left(M_{\rm BH}/M_\odot\right) = M_{\rm BH} - \alpha_{\rm local}\log\left(M_*/M_\odot\right)-\beta_{\rm local},
\end{equation}
For $\alpha_{\rm local}$ and $\beta_{\rm local}$, we use the above values from the local samples \citep{Haring2004, Bennert2011}, i.e., $0.97$ and $-2.48$, respectively. 
We plot $\Delta\log\left(M_{\rm BH}/M_\odot\right)$ as a function of $z$ in the right panel of Figure~\ref{fig:massmass}. We then parameterize $\Delta\log\left(M_{\rm BH}/M_\odot\right)$ to evolve with $z$ as:
\begin{equation}
    \Delta\log\left(M_{\rm BH}/M_\odot\right) = \gamma\log\left(1+z\right). \label{eq:Delta}
\end{equation}
Here, we assume there is no redshift change in $\alpha$ and $\beta$ and $\Delta\log M_{\rm BH}$ can be described by only the evolution from the local relation ($\log M_{\rm BH} = \alpha_{\rm local} \log M_{\rm *} + \beta_{\rm local}$).
Fitting our data with Equation~(\ref{eq:Delta}) without considering the selection bias results in $\gamma=1.33^{+0.13}_{-0.14}$, suggesting positive evolution.
Although $R^2$ for this fitting is 0.050, the BIC for this fitting is significantly lower than the $\gamma=0$ model (i.e., $\Delta\log\left(M_{\rm BH}/M_\odot\right) = 0$ with a scatter); thus, this model describes the data better than the $\gamma=0$ model.
However, because our sample is (X-ray) flux-limited, it's essential to consider the impact of selection bias when determining the mass relation \citep[e.g.,]{Ding2020_HST, Li2021_HSC, Li2021_EoR}.
Thus, for the rest of this section, we measure the intrinsic slope of the mass relation ($\alpha$) at $z\sim1.5$ and the redshift evolution parameter ($\gamma$) by comparing our results with mock catalogs as described in Section~\ref{subsubsec:_mock}. This approach allows us to account for selection bias and measurement uncertainties to determine the intrinsic redshift evolution and dispersion of the mass relation.

\begin{figure*}[ht!]\epsscale{1.15}
  \plotone{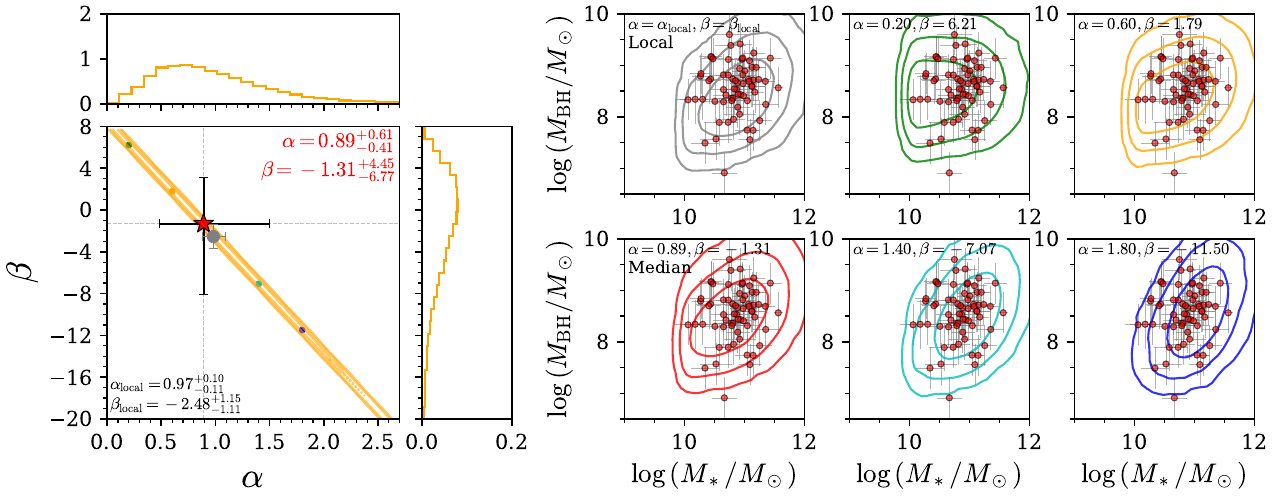}
  \caption{
  ({\it left}) Posterior distribution in the $\alpha-\beta$ plane obtained through a comparison between mock observation and our sample. 
  The posterior distributions for $\alpha$ and $\beta$ are shown in the right and upper 1D histogram.
  Each contour indicates $1$, $2$, and $3\sigma$ levels.
  The median and $1\sigma$ confidence levels of $\alpha$ and $\beta$ are shown in the top-right corner of each panel.
  ({\it right}) Comparison of mock observed masses generated with each $\alpha$ and $\beta$ at $z=1.5$ with the real observed data at $z=1-2$ shown in red circles.
  Contours show the distribution of mock observation, indicating $1$, $2$, and $3\sigma$ levels.
  Each parameter set, $\alpha$ and $\beta$, is shown in the upper left corner of each panel.
  As shown, we cannot reject the scenario that $\alpha$ does not evolve up to $z\sim2.5$.
  \label{fig:alpha}}
\end{figure*}

\subsection{Intrinsic slope ($\alpha$) of the mass relation at $z\sim1.5$}\label{subsec:sel_alpha}

Past efforts to establish the evolution of the ratio between black hole and galaxy mass assume that there is a linear relation at higher redshifts and the slope of the relation matches the local relation \citep[e.g.,][]{Ding2020_HST, Li2021_HSC}. 
This may not necessarily be the case. 
Here, we assess how well the parameters of a linear relation can be constrained.  

In particular, we initially assumed that $\alpha$ and $\beta$, for constructing the mock samples (Equation~(\ref{eq:mock_mr}), have constant values independent of the redshift, and the evolution is expressed simply as $\gamma\log\left(1+z\right)$. However, the results in the previous Section~\ref{subsec:mass-mass} may indicate a different $\alpha$ from the local value without considering selection biases.
In this section, we examine whether the constant $\alpha$ assumption is valid by estimating the intrinsic value of $\alpha$ while considering selection bias and provide evidence for an intrinsic relation between $M_{\rm BH}$ and $M_*$ at $z > 1$ for the first time.

Since the parameters $\alpha$, $\beta$, and $\gamma$ exhibit degeneracies, in this subsection, we constrain the $z$ range to $z=1-2$ and perform fitting for $\alpha$ and $\beta$ with an assumption of $\gamma=0$ over this redshift range.
Consequently, we cannot compare the estimated $\beta$ with the values in the local relation directly.
For the intrinsic dispersion of the mass relation $\sigma_\mu$, we assume 0.3 dex in this subsection, thus matching local studies.

\begin{figure*}[ht!]\epsscale{1.15}
  \plotone{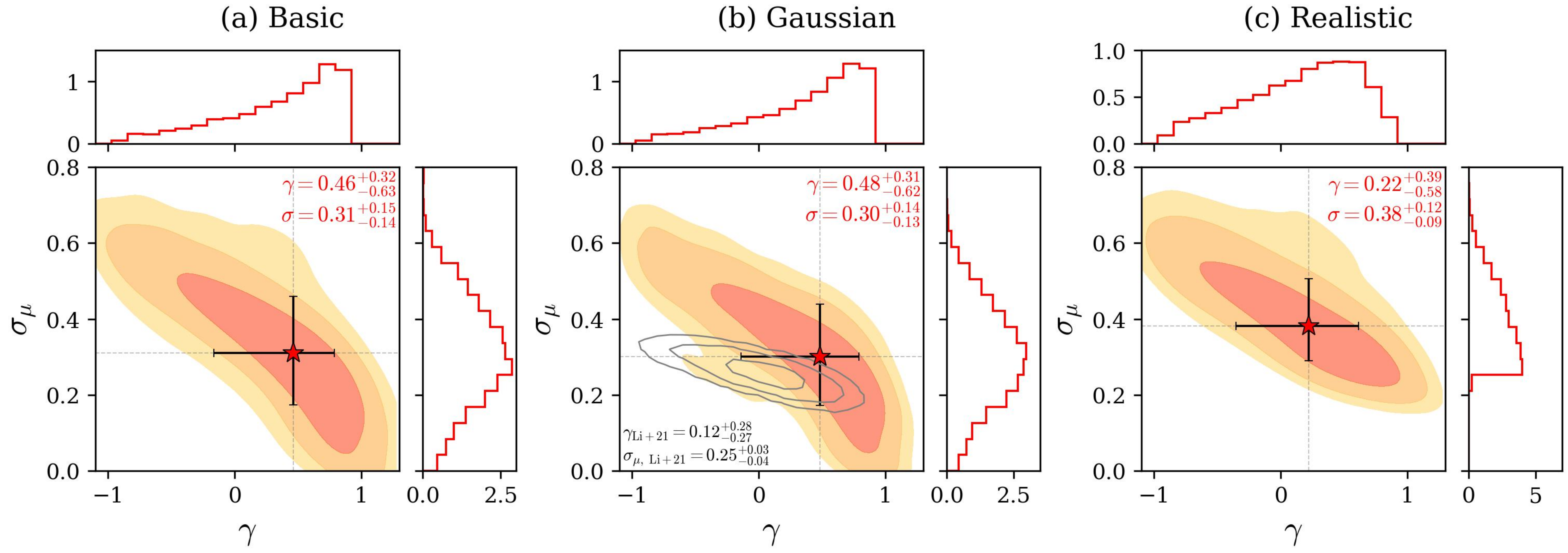}
  \caption{Posterior distribution on the $\sigma_\mu-\gamma$ plane obtained through a comparison between mock observation and our sample (the result with the top-5 stacked PSFs).
  Panels (a) to (c) correspond to the result using basic, Gaussian, and realistic prior settings, respectively.
  The posterior distribution for $\sigma_\mu$ and $\gamma$ are shown in the right and upper 1D histogram.
  Each contour (red to orange) indicates $1$, $2$, and $3\sigma$ from inner to outer.
  The median and $1\sigma$ confidence levels of $\gamma$ and $\sigma_\mu$ are shown in the top-right corner of each panel.
  Only in panel~(b), we plot the contour from \cite{Li2021_HSC} in gray, and their $\gamma$ and $\sigma_\mu$ estimations are shown in the lower left corner based on SDSS quasars at $0.3<z<0.8$ with Subaru HSC imaging.
  Regardless of the prior, the results suggest no or mild evolution at $z < 2.5$.
  \label{fig:selbias}}
\end{figure*}

With the mock observed data described in Section~\ref{subsubsec:_mock}, we apply the selection criteria corresponding to our sample (Section~\ref{subsec:sample_selection}) and compare them with the observed results to constrain the evolution parameters.
We assume the observation thresholds as;
\begin{align}
    F_{\rm [2-10~keV]} &> F_{\rm [2-10~keV], lim},\\
    FWHM_{\rm line} &> 1000~{\rm km~s^{-1}},\\
    M_* &> 10^{10}M_\odot
\end{align}
where the first and second conditions correspond to the detection limit of the X-ray observation and broad lines for single epoch $M_{\rm BH}$ estimation.
The third condition accounts for the detection limit of the host galaxies, and we confirm that galaxies with masses greater than $10^{10}M_\odot$ are successfully detected across all redshift ranges in our decomposition analysis.
Thus, in comparing mock and real observations, galaxies with masses below $10^{10}M_\odot$ are excluded to maintain consistency with this mock observation.
As shown in Figure~\ref{fig:sample_range} (a), excluding the eight objects, all sources in our sample have $F_{\rm [2-10~keV]}$ above the XMM-Newton $F_{\rm [2-10~keV], lim}$.
Note that all of the eight targets with $F_{\rm [2-10~keV]}$ smaller than XMM-Newton $F_{\rm [2-10~keV], lim}$ are targets observed only with Chandra. 
We changed $F_{\rm [2-10~keV], lim}$ depending on in which survey each target was detected.
Because $M_{\rm BH}$ of our real targets are based on single epoch estimation with three different lines (${\rm H\alpha}$, ${\rm H\beta}$, and ${\rm Mg{\sc II}}$), we have three different selection thresholds on the mock galaxy using $FWHM_{\rm H\alpha}$, $FWHM_{\rm H\beta}$, and $FWHM_{\rm Mg{\sc II}}$.

For each of our targets, we first select the mock galaxies with the corresponding selection bias, i.e., the selection condition using the same line information used to estimate $M_{\rm BH}$. 
Furthermore, we select the mock galaxies with a similar redshift; $\left|z-z_{\rm mock}\right|<0.1$.
Then, we calculate the probability that the mock galaxies with the similar $z_{\rm mock}$ as the real galaxy would have the same $\Tilde{M}_{\rm *,mock}$ with $\left|\Delta M_{\rm *}\right|<0.1$ and $M_{\rm vir,mock}$ with $\left|\Delta M_{\rm BH}\right|<0.1$, i.e., calculate the probability $p$ following
\begin{equation}
    p = \frac{N_{\left|\Delta z\right|<0.1,~\left|\Delta M_*\right|<0.1,~\left|\Delta M_{\rm BH}\right|<0.1}}{N_{\left|\Delta z\right|<0.1}}.
\end{equation}
We then calculate the likelihood of our sample being observed for each parameter combination of $\alpha$ and $\beta$. 
Thereby, we estimate the probability distribution of these parameters using MCMC.
In the sampling, we assume a uniform prior between 0.1 to 3 for $\alpha$, and -25 to 8 for $\beta$.

\begin{figure*}\epsscale{1.0}
  \plotone{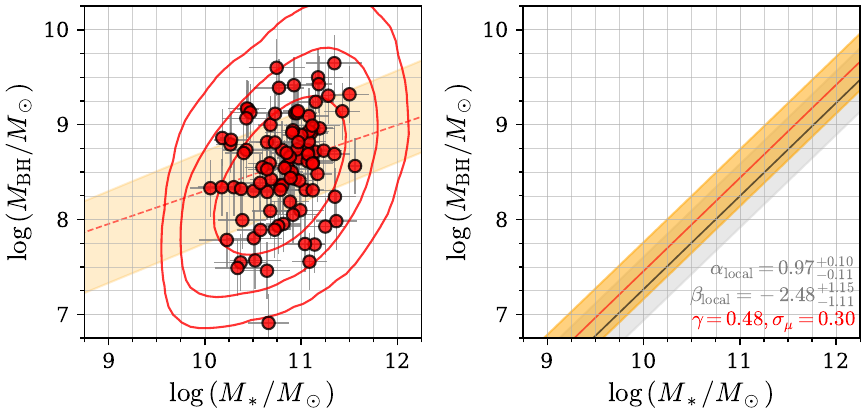}
  \caption{
  ({\it left}) Comparison of the observed sample and the mock observation of $z\sim1.5$ AGN on the $M_{\rm BH}-M_*$ plane.
  The mock data at $z\sim1.5$ is constructed with the best-fit $\gamma$ and $\sigma_\mu$ (the result with the top-5 stacked PSFs assuming the ``Gaussian'' prior) while assuming $\alpha_{\rm local}$ and $\beta_{\rm local}$.
  Small red circles with error bars and the red dashed lines indicate the observed data and the linear fit results (same with Figure~\ref{fig:massmass} {\it left}).
  Contours indicate a distribution of mock observed samples, showing $1$, $2$, and $3\sigma$ from inner to outer.
  ({\it right})
  Comparison of the intrinsic $M_{\rm BH}$--$M_*$ relation at $z\sim1.5$ (the red line) to the local relation (the gray line).
  \label{fig:massmass_final}}
\end{figure*}

The obtained $\alpha-\beta$ distribution is shown on the left panel of Figure~\ref{fig:alpha}, indicating a strong anti-correlated degeneracy between $\alpha$ and $\beta$. 
Because the $1\sigma$ contour includes values $\left(\alpha_{\rm local}, \beta_{\rm local}\right)$ corresponding to the local relation, our results do not definitively reject the scenario where $\alpha$ and $\beta$ do not evolve compared to the local relation up to $z\sim2.5$.
In the right panels of Figure~\ref{fig:alpha}, we show the $M_{\rm BH}-M_*$ distribution for mock observed galaxies (similar to Figure~\ref{fig:mockmake}f) generated with manually sampled parameters on the ridge of the $\alpha-\beta$ degeneracy and the real observed galaxies ($\alpha\sim0.2-1.8$).
The plots suggest that the mock data exhibits a similar distribution to our sample within the comparable mass range.
To break this degeneracy and improve the precision of determining $\alpha$ and $\beta$, a larger sample with a wider $M_{\rm BH}$ range in future studies is essential.
Even so, we demonstrate that a relation between $M_{\rm BH}$ and $M_*$ at high-$z$ is realized based on having a statistical sample afforded by the COSMOS-Web data set.
As indicated by Figure~\ref{fig:comp_res_deco} (d), the difference in the PSF reconstruction methods does not significantly affect the $M_*$ estimation. Therefore, the posterior distribution of $\alpha$ and $\beta$ that are estimated based on $M_*$ also shows no significant PSF dependency.

\subsection{Intrinsic evolution ($\gamma$) of the $M_{\rm BH}/M_*$ relation}\label{subsec:redshift_evolution}
To determine the evolution of mass relation with consideration of the selection bias and measurement uncertainties, we generate mock observed catalogs with free parameters of $\gamma$ (evolution rate) and $\sigma_\mu$ (intrinsic dispersion of the mass relation) and constrain them by comparing the mock catalogs with observational data in a similar manner to Section~\ref{subsec:sel_alpha}.
Note that the assumption of $\gamma=0$ and $\sigma_\mu=0.3$ used in section~\ref{subsec:sel_alpha} is not applied in the fitting performed in this subsection; both parameters are treated as free parameters during the fitting process here.
In contrast to Section~\ref{subsec:sel_alpha}, we fixed $\alpha$ and $\beta$ in Equation~(\ref{eq:mock_mr}) to the values in the local relation ($\alpha_{\rm local}$, $\beta_{\rm local}$) to consider the redshift evolution.
As illustrated in Section~\ref{subsec:sel_alpha}, we cannot rule out the possibility for evolution of $\alpha$.
However, as described in Section~\ref{subsec:sel_alpha}, $\alpha$, $\beta$, and $\gamma$ are strongly degenerate, and obtaining physically meaningful results is challenging when all three parameters are left free for the sample being considered here.

There is still a possibility that $\sigma_\mu$ depends on redshift. Nevertheless, as discussed later, imposing strong constraints on $\sigma_\mu$ in our results is challenging due to the sample size and its uncertainties.
Therefore, we set $\sigma_\mu$ as a constant independent of redshift in the fitting.
It means $\sigma_\mu$ obtained through this method is considered to be an averaged value over $z\sim0.68-2.5$.
Even so, this $\sigma_\mu$ estimation has the highest statistical significance for such a study at $z \gtrsim 1$.
We discuss the redshift evolution of $\sigma_\mu$ in Section~\ref{subsubsec:scatter_evol}.
Finally, in this analysis, we assume no redshift-dependent parameters among the free parameters.
Therefore, there is no need to restrict the redshift range within the data, as in Section~\ref{subsec:sel_alpha}.

In the fitting process, we assume a uniform prior distribution for $\gamma$ between $-1<\gamma<1$.
Then, we have three different prior settings for $\sigma_\mu$: a uniform distribution between $0.01<\sigma_\mu<1.0$ (basic), a Gaussian distribution with a mean of 0.3~dex and a standard deviation of 0.1~dex (Gaussian), and a uniform distribution between $0.25<\sigma_\mu<1.0$ with a prohibition of $\sigma_\mu<0.25$ (realistic).

Figure~\ref{fig:selbias} shows the estimated posterior distributions of $\gamma$ and $\sigma_\mu$ using each prior setting with the top-5 stacked PSF results. 
As evident in all panels by the orange contours, the intrinsic dispersion $\sigma_\mu$ is strongly degenerate with the evolution rate $\gamma$ where a smaller $\gamma$ results in a larger $\sigma_\mu$ as demonstrated in previous studies \citep{Ding2020_HST, Li2021_HSC}. 
To reiterate, a smaller value for $\gamma$ biases the mass relation towards relatively lower $M_{\rm BH}$ values thus a larger $\sigma_\mu$ is required to reproduce a certain set of observation data.

Considering the likelihood distribution for $\gamma$, the ``basic'' prior setting shows a slight positive-to-no evolution with $\gamma=0.46^{+0.32}_{-0.63}$. 
Similarly, the  ``Gaussian'' prior setting results in $\gamma=0.48^{+0.31}_{-0.62}$.
If we assume that the intrinsic dispersion should not be significantly smaller than the local dispersion, we limit the allowed range for $\sigma_\mu$ to be above $0.25$ ("Realistic" case). 
In this case, we find $\gamma=0.22^{+0.39}_{-0.58}$, closer to the case for no evolution ($\gamma=0$) than the results with the other priors.
For the latter, the intrinsic dispersion is slightly higher at $0.38^{+0.12}_{-0.09}$. 
In all cases, our results are consistent with very mild or essentially a lack of evolution with respect to the local relation.

Then, the left panel of Figure~\ref{fig:massmass_final} compares mock observations using the median parameters ($\gamma=0.48$ and $\sigma_\mu=0.30$) under the assumption of ``Gaussian'' prior with the actual observational $M_{\rm BH}$--$M_*$ distribution and relation.
We can see that the mock data can explain the observed data well. 
The right panel of Figure~\ref{fig:massmass_final} compares the intrinsic relationship, i.e., the relation corrected for the selection bias, with the local relation.
Again, the resulting intrinsic relation is consistent with the local relation within the range of errors.


\section{Discussion}\label{sec:discussion}
In Section~\ref{subsec:redshift_evolution}, our findings suggest a mild or lack of evolution of the mass relation from the local relation when considering selection biases and measurement uncertainties. 
In this section, we first compare the derived values of $\gamma$ and $\sigma_\mu$ from Section~\ref{subsec:redshift_evolution} with previous studies.
Then, we also discuss the cosmic averaging scenario \citep{Peng2007,Hirschmann2010,Jahnke2011}.

\subsection{Comparison to other studies}
First, the conclusion of no or mild evolution with $\gamma=0.48^{+0.31}_{-0.62}$ is consistent with studies based on 2D decomposition analysis \citep[e.g.,][]{Ding2020_HST, Li2021_HSC} and studies using a SED-fitting-based decomposition method \citep[e.g.,][]{Sun2015, Suh2020}.

In Figure~\ref{fig:selbias}, we compare our results on the estimated $\gamma-\sigma_\mu$ distribution to those of \cite{Li2021_HSC}.
Our sample has higher redshift range than \cite{Li2021_HSC}, and the change of $\Delta\log\left(M_{\rm BH}/M_\odot\right)$ is also proportional to $\log\left(1+z\right)$ in our model.
Therefore, to reproduce the observational results, when increasing $\gamma$, we need to decrease $\sigma_\mu$ more than \cite{Li2021_HSC}.
In other words, the slope of the $\gamma$--$\sigma_\mu$ degenerate relation is steeper in our study.
As a result, while the sample size of this study is approximately six times smaller than \cite{Li2021_HSC}, the uncertainty of $\gamma$ is only $\sim2$ times larger than \cite{Li2021_HSC}.
On the other hand, due to the steep slope, imposing constraints on $\sigma_\mu$ becomes challenging, and the uncertainty becomes $\sim4$ times larger than \cite{Li2021_HSC}.
Even so, our estimated value of $\sigma_\mu$ is $0.30^{+0.14}_{-0.13}$, which is remarkably similar to \cite{Li2021_HSC} with $\sigma_\mu$=$0.25_{-0.04}^{+0.03}$.

It is worth highlighting that the inference on the value of $\gamma$ is very close to zero for the "Realistic" case (Fig.~\ref{fig:selbias}c) where we assume that the intrinsic scatter ($\sigma_\mu$) cannot be lower than the local dispersion. 
Interestingly, if $\sigma_\mu$ ($\sim0.4-0.5$) is actually higher than the local value, this would push the evolution parameter to negative values ($\gamma\sim-0.5$), thus presenting a scenario where the black holes have to catch up to their host galaxies by a bit.

\subsection{Scatter ($\sigma_\mu$) evolution and cosmic averaging}\label{subsubsec:scatter_evol}

When assuming a non-casual cosmic averaging scenario \citep{Jahnke2011}, major mergers average and equalize the mass ratio $M_{\rm BH}/M_*$ through cosmic history. Thus, $\sigma_\mu$ should increase towards high redshift.

To test the cosmic averaging scenario with our data, we generate a mock sample (20,000 parameter sets of $M_{\rm BH}$ and $M_*$) at $z\sim1.4$, the median redshift of our observational sample.
We generate $M_*$ based on the SMF by \cite{Weaver2023}, and calculate the $M_{\rm BH}$ assuming the $\gamma$ and $\sigma$ sampled in the MCMC run with ``Gaussian'' prior setting (Section~\ref{subsec:redshift_evolution}). 
Each mock galaxy is assumed to undergo major mergers following the major merger rate from \cite{Rodriguez2015} covering $z\sim0-1.4$. We simulate the redshift evolution of $M_{\rm BH}$ and $M_*$ by summing them with those of merging partners.
Then, we calculate $\sigma_\mu$ at each redshift to trace the expected scatter evolution from the assumed conditions.
In this simulation, we do not consider the accretion onto the black hole and star formation, i.e., both $M_{\rm BH}$ and $M_*$ are assumed to grow only through major mergers.
We limit the mergers to those with mass ratios within $\pm 0.5~{\rm dex}$.

\begin{figure}[ht!]\epsscale{1.15}
  \plotone{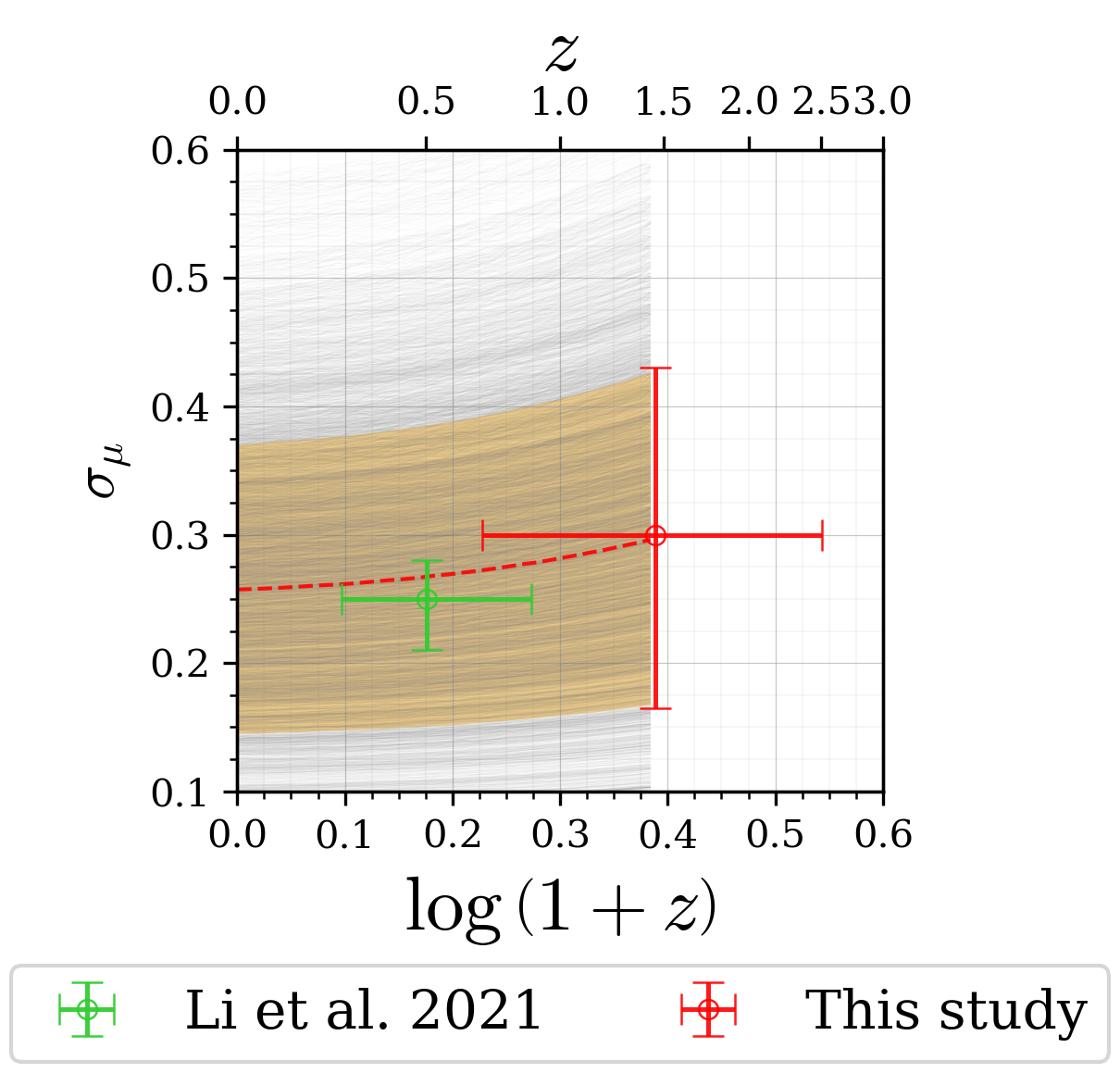}
  \caption{Evolution of $\sigma_\mu$ at $z\lesssim2$ based on a simple simulation of the cosmic averaging scenario.
  The vertical red error bars represent the scatter in $\sigma_\mu$ within the $z$ range of our sample using the basic prior. Faint grey lines correspond to each sampled parameter set while the red line and the orange filled region represent the median and 1$\sigma$ confidence level of $\sigma_\mu$ redshift evolution.
  The green data denotes the constraint from \cite{Li2021_HSC} which is consistent within the $1\sigma$ range of our results.
  \label{fig:scatter_evol}}
\end{figure}

Figure~\ref{fig:scatter_evol} compares the simulated redshift evolution of $\sigma_\mu$ with the results from this study and \cite{Li2021_HSC}.
When assuming the evolution only through major mergers, the growth within the $1\sigma$ uncertainty range significantly encompasses the results of \cite{Li2021_HSC}.
Moreover, our median redshift evolution is consistent with the results of \cite{Li2021_HSC}.
Therefore, the $\sigma_\mu$ difference between the results from this study and \cite{Li2021_HSC} could be interpreted as the major merger-based scatter evolution.

However, due to the large uncertainty in our results, we cannot draw any definitive conclusions regarding the redshift evolution of $\sigma_\mu$.
Furthermore, as evident from Figure~\ref{fig:scatter_evol}, our sample has a wide redshift range compared to \cite{Li2021_HSC}.
If $\sigma_\mu$ varies with redshift, the sample should be binned in a narrower redshift range to trace the redshift evolution.
Nevertheless, as mentioned in Sections~\ref{subsec:sel_alpha} and \ref{subsec:redshift_evolution}, the degeneracy relation on the $\gamma-\sigma_\mu$ plane tends to steepen toward higher redshifts, making it relatively challenging to impose constraints on $\sigma_\mu$. Future high-$z$ statistical studies will likely require samples of a similar size to \cite{Li2021_HSC} with $N=584$, or even larger to address the redshift evolution of $\sigma_\mu$. Thus, it will be necessary to conduct comprehensive surveys of high-$z$ AGNs using next-generation survey data such as Euclid and Roman.

\subsection{General notes on PSF reconstruction methods}\label{subsec:PSF_dependence}
So far, various studies have performed decomposition of JWST images \citep[e.g.][]{Ding2022_CEERS, Ding2022_z6, Stone2023, Yue2023, Harikane2023, Zhuang2023, Zhuang2023COSMOS, Stone2023undermassive}.
As discussed above or in the previous studies, the results of the AGN+host galaxy 2D decomposition depend significantly on the PSF reconstruction.
Especially, \cite{Zhuang2023} discussed the effect of different PSF on decomposition results.
\cite{Zhuang2023} compared three PSF modeling methods ({\tt SWarp}, {\tt photutils}, and {\tt PSFEx}), but they did not compare them with $\chi^2_\nu$ based methods directly.
Thus, in this paper, we summarize the comparison when using different PSF reconstruction methods.

In this study, we compare three final PSFs: two obtained by $\chi^2_\nu$-based methods (a top-5 stacked PSF and a top-75\% stacked PSF) and one from an empirically modeling method, {\tt PSFEx}. 
As demonstrated in Figure~\ref{fig:fwhm_comp}, we find offsets in FWHM among different PSF reconstruction methods. 
These PSF variations, as indicated by \cite{Zhuang2023}, could potentially introduce biases in estimating $n$, $r_e$, or $H/T$. This is due to a broader (narrower) PSF than in reality which tends to result in smaller (larger) $n$, larger (smaller) $r_e$, and overestimation of host fluxes.
However, as shown in Figure~\ref{fig:comp_res_deco}, morphological parameters such as $n$, $r_e$, $H/T$ generally exhibit consistent relations.
Besides, as shown in Figure~\ref{fig:comp_res_deco} (d), different PSFs have less impact on $M_*$ estimation than on $H/T$.
This could be due to the fact that $M_*$ is estimated from SED fitting (Section~{\ref{subsec:sed_fitting}}) using multi-band photometry that averages the uncertainty in each band.
If so, SED fitting with a smaller number of photometric bands \citep[e.g.,][]{Ding2022_z6, Yue2023}, may lead to more severe effects from inaccurate PSF reconstruction. We also find each method has advantages and disadvantages from a technical aspect. Lastly, we summarize below the technical comparison between each method.

{\bf \underline{Modeling method:}}
The approach of constructing an empirical PSF model from numerous stars, such as {\tt PSFEx} in this study, has the advantage of being less influenced by noise compared to the $\chi^2_\nu$-based methods, as depicted in Figure~\ref{fig:final_PSF}.
This method also allows flexible modeling, considering PSF as a function of position or brightness.

However, a drawback is the requirement for many PSF candidates to model a local or flux-dependent PSF, which can be considered a trade-off.
Furthermore, we also confirm that FWHM values of {\tt PSFEx} PSFs depend on configuration parameters.
When reconstructing PSFs, it is challenging to determine the best configuration parameters because the exact PSFs are not known.

{\bf \underline{$\chi^2_\nu$-based selection method:}}
Selecting PSFs based on $\chi^2_\nu$ from a substantial number of stars, such as Top-5 or Top-75\% PSFs, allows easy analysis considering the PSF uncertainties. 
Also, by fitting with various different single PSFs, the possibility for PSF mismatches is minimized. 
Notably, \cite{Yue2023} discussed the differences in broad-band PSFs attributed to variations in SED shapes between stars and AGNs.
Our method using all single PSFs in the PSF library may consider this PSF uncertainty as a result of selecting low-$\chi^2_\nu$ single PSFs with a matched shape.

However, using fewer PSF candidates, like the top-5 stacked PSF, might increase the noise of the final PSF, as observed in Figure~\ref{fig:final_PSF}. 
Note that this noise is generally smaller than the central main component; thus, it should not affect significantly except in the case with small $H/T$. 
Additionally, our approach involves visual inspection in PSF candidate selection, which might cause a bias.
Also, as mentioned by \cite{Zhuang2023}, a lower $\chi^2_\nu$ does not necessarily mean more correct PSF.
Moreover, as a practical demerit, this method needs more time to create a PSF library with visual inspection in each filter and region and more computational cost for SED fittings with all single PSFs.

Given that we do not know the correct answers in this study, it is challenging to discuss which method produces the most accurate results. 
For some targets, decomposition is clearly successful with one final PSF, which can then be evaluated from the residual emission based on other PSFs.
We also confirm that these failures in fitting can occur with all final PSFs.
\textit{Therefore, we conclude that it is best to assess the impact on derived properties (e.g., $n$, $r_e$, $H/T$) by varying the method of PSF construction (Section~\ref{subsec:psf}) and place equal weight on assessing the uncertainties based on varying PSFs (Section~\ref{subsubsec:detection}) to obtain solid 2D decomposition results}.


\section{Conclusions} \label{sec:conclusion}
We performed a 2D decomposition analysis of \hbox{high-$z$} ($z\sim0.68-2.5$) type-I AGNs using the COSMOS-Web \citep{Casey2022} and PRIMER-COSMOS surveys to measure the black hole -- stellar mass relation at high-$z$.
Our sample contains 107 targets that are X-ray-selected, broad-line AGNs with single-epoch black hole mass estimates ($\log\left(M_{\rm BH}/M_\odot\right)\sim6.9-9.6$) based on ${\rm H}\beta$, ${\rm H}\alpha$, and Mg{\sc ii} from previous spectroscopic surveys \citep[e.g.,][]{Schulze2018}.

By utilizing HST/ACS + JWST/NIRCam imaging that covers the rest-frame optical to near-infrared, we obtained multi-band information of the AGN host galaxies with unprecedented spatial resolution in which we can clearly identify substructures such as bars and spirals arms (Figure~\ref{fig:3col}).
Since AGN--host 2D decomposition is known to be sensitive to the PSF reconstruction methods, we compared the results with three final PSFs reconstructed using the modeling method {\tt PSFEx} and a $\chi^2_\nu$-based selection method. 
Through a meticulous decomposition analysis using various PSFs, we successfully detected the host galaxies in more than two filters for over $90\%$ of the entire sample.
Then, we confirmed that host morphological parameters such as $n$, $r_e$, and $H/T$ remain relatively consistent regardless of the PSF reconstruction method used (Figure~\ref{fig:comp_res_deco}).
Furthermore, given the high quality of the host galaxy images, this study is expected to serve as a crucial stepping stone for image-based spatially-resolved investigations of AGN host galaxies, such as double-\sersic model fitting (decomposition fitting with AGN, bulge, and disk components) or parametric/non-parametric substructure analysis.

With AGN-subtracted photometry of the host galaxy in multiple bands, we estimate $M_*$ by performing SED fitting and present the $M_{\rm BH}$--$M_*$ relation at $z\sim0.68-2.5$ (Figure~\ref{fig:massmass}).
There is a weak positive correlation between $M_{\rm BH}$ and $M_*$ with the correlation coefficient of $\rho=0.25$ ($p=0.010$).
We fit the mass relation by a simple \hbox{(log-)linear} relation of $\log\left(M_{\rm BH}/M_\odot\right) = \alpha\log\left(M_*/M_\odot\right) + \beta$ with consideration of selection biases and measurement uncertainties.
Our results show that the slope of the mass relation at $z\sim2$ ($\alpha=0.89^{+0.61}_{-0.41}$) is consistent with the local relation ($\alpha_{\rm local}=0.97^{+0.10}_{-0.11}$) (Figure~\ref{fig:alpha}).

Assuming the redshift evolution term of the mass relation to be $\gamma\log\left(1+z\right)$, we further determine the evolution factor $\gamma$ and the intrinsic scatter of the mass relation $\sigma_\mu$ while considering selection biases and uncertainties based comparisons to mock catalogs (Figure~\ref{fig:selbias}).
Even though the estimated probability distribution shows strong degeneracy between $\gamma$ and $\sigma_\mu$, we find no or mild evolution with $\gamma=0.48^{+0.31}_{-0.62}$.
If we assume that $\sigma_\mu$ is not smaller than the local value, we obtain $\gamma=0.22^{+0.39}_{-0.58}$, which is more consistent with the no- or mild-evolution scenario.
The estimated $\gamma-\sigma_\mu$ distribution is largely consistent with \cite{Li2021_HSC} based on the HSC imaging of SDSS quasars at $z<0.8$.
Given the higher redshift range of our sample, the slope of the degeneracy relation between $\gamma$ and $\sigma_\mu$ is steeper than \cite{Li2021_HSC}. 
Therefore, despite the sample being approximately six times smaller, the estimated $\gamma$ uncertainty is just slightly larger than \cite{Li2021_HSC}.

Furthermore, the estimated value of the intrinsic scatter is $\sigma_\mu=0.30^{+0.14}_{-0.13}$ which is consistent with the local relation and the recent estimate by \citet{Li2021_HSC}.
We show that this value $\sigma_\mu$ at high-z may not be in contradiction to a cosmic averaging scenario (Figure~\ref{fig:scatter_evol}) as recently put forward by \citet{Li2021_HSC} and \citet{Ding2022_modelcomp} where AGN feedback is invoked to explain the constant level of dispersion with redshift.
However, due to the small sample size, high redshift, and wide redshift range, our constraints on the redshift evolution of $\sigma_\mu$ are weak.
Thus, a larger sample size at $z\sim1-3$ is needed, especially at high-$z$.

Future large-scale surveys such as Euclid and Roman will significantly augment the sample size, along with deeper observations by JWST, to provide stronger constraints on SMBH and galaxy evolution. For future large imaging data sets, visual inspection for all multi-component fits and manually exclusion of anomalous results will not be feasible; thus, improvements in 2D decomposition techniques or the imposition of more sophisticated conditions to confirm the robustness of host detection is needed.
Additionally, by leveraging the high spatial resolution of JWST images, it is important to compare the morphology, substructures (shortly introduced in Section~\ref{subsec:hostimage}), and stellar populations of AGN-host galaxies with non-AGN galaxies, as well as to discuss the presence or absence of a bulge component and the $M_{\rm BH}-M_{\rm bulge}$ relation.
In order to address these challenges, it is imperative to enhance the 2D decomposition analysis by mitigating the uncertainties related to PSF reconstruction. 
This involves conducting a meticulous analysis of AGN PSFs, determining the validity of applying stellar PSFs to AGN with different SEDs than stars, identifying the most effective methods for accurate PSF reconstruction, and establishing a framework for evaluating the uncertainties in reconstructed PSFs. 


\begin{acknowledgments}
This work is based on observations made with the NASA/ESA/CSA James Webb Space Telescope.
The data were obtained from the Mikulski Archive for Space Telescopes at the Space Telescope Science Institute, which is operated by the Association of Universities for Research in Astronomy, Inc., under NASA contract NAS 5-03127 for JWST.
These observations are associated with program IDs 1727 and 1837, and the specific observations analyzed can be accessed via \dataset[DOI]{https://doi.org/10.17909/g0bh-7854}.
Numerical computations were in part carried out on the iDark cluster, Kavli IPMU.
This work was made possible by utilizing the CANDIDE cluster at the Institut d’Astrophysique de Paris, which was funded through grants from the PNCG, CNES, DIM-ACAV, and the Cosmic Dawn Center and maintained by Stephane Rouberol.
We thank Marko Shuntov, Kei Ito, and Mingyang Zhuang for giving us useful advice for reconstructing the PSFs.
We thank Mingyang Zhuang, Yue Shen, and Junyao Li for giving us their measurements for comparison.
We thank the anonymous referee for helpful feedback.
Kavli IPMU is supported by World Premier International Research Center Initiative (WPI), MEXT, Japan.
The Cosmic Dawn Center (DAWN) is funded by the Danish National Research Foundation under grant DNRF140.
TT is supported by Forefront Physics and Mathematics Program to Drive Transformation (FoPM), a World-leading Innovative Graduate Study (WINGS) Program, the University of Tokyo.
JS is supported by JSPS KAKENHI (JP22H01262) and the World Premier International Research Center Initiative (WPI), MEXT, Japan.
This work was supported by JSPS Core-to-Core Program (grant number: JPJSCCA20210003).
BT acknowledges support from the European Research Council (ERC) under the European Union’s Horizon 2020 research and innovation program (grant agreement number 950533) and from the Israel Science Foundation (grant number 1849/19).
JR is supported by JPL, which is run by Caltech for NASA.
GEM and SG acknowledge the Villum Fonden research grant 13160 “Gas to stars, stars to dust: tracing star formation across cosmic time,” grant 37440, “The Hidden Cosmos,” and the Cosmic Dawn Center of Excellence funded by the Danish National Research Foundation under the grant No. 140. 

\end{acknowledgments}

%

\vspace{5mm}
\facilities{JWST (NIRCam), HST (ACS)}


\software{astropy \citep{Astropy2013,Astropy2018,Astropy2022},
          CIGALE \citep{Boquien2019, Yang2022},
          FSPS \citep{Conroy2009, Conroy2010},
          galight \citep{Ding2020_HST},
          lenstronomy \citep{2018PDU....22..189B, 2021JOSS....6.3283B},
          matplotlib \citep{Hunter2007_matplotlib},
          wCorr \citep{wCorr},
          numpy \citep{harris2020_numpy},
          photutils \citep{2023zndo...7946442B},
          psfr (Birrer et al. in prep),
          Prospector \citep{Leja2017,Johnson2021},
          jwst \citep{jwst_pipeline}
          }



\appendix

\section{Detected numbers in each filter}\label{app:detnum}
Table~\ref{tab:detection} summarizes the number of undetected host galaxies based on the conditions described in Section~\ref{subsubsec:detection} for each PSF and filter, while Table~\ref{tab:detnum} then summarizes the number of detected hosts for each final PSF. Regarding the JWST filters, we have less galaxies that are classified as non-detection due to the ${\rm BIC}$ condition than due to the $S/N_{\rm host}$ condition.
The number of undetected hosts is lowest in F277W, followed by F150W and F444W.
On the other hand, F115W and F814W show a larger number of undetected hosts, especially due to $S/N_{\rm host}$. 
This trend is because F814W is the HST observation with lower resolution and shallower depth than the JWST observations, and both F814W and F115W are in the shorter-wavelength side of the Balmer break, leading to a tendency for a smaller intrinsic $H/T$. 

\begin{table}[]
\caption{The number of the undetected host galaxies for each condition, PSF and filter}\label{tab:detection}
\begin{tabular}{llllll}
\hline\hline
 & & \multicolumn{1}{c}{(1)} & \multicolumn{1}{c}{(2)} & \multicolumn{1}{c}{(3)} & \\
Filter & PSF & BIC & $S/N$ & manual & Total\\ \hline
\multirow{3}{*}{F814W} & Top-5 & 48 & 53 & 30 & 90\\
& Top-75\% & 51 & 61 & 29 & 94\\
& PSFEx & 53 & 58 & 29 & 94\\ \hline
\multirow{3}{*}{F115W} & Top-5 & 1 & 20 & 1 & 21\\
& Top-75\% & 1 & 21 & 1 & 22\\
& PSFEx & 2 & 21 & 1 & 22\\ \hline
\multirow{3}{*}{F150W} & Top-5 & 0 & 10 & 1 & 11\\
& Top-75\% & 0 & 9 & 1 & 10\\
& PSFEx & 0 & 8 & 1 & 9\\ \hline
\multirow{3}{*}{F277W} & Top-5 & 0 & 1 & 1 & 2\\
& Top-75\% & 0 & 0 & 1 & 1\\
& PSFEx & 0 & 1 & 1 & 2\\ \hline
\multirow{3}{*}{F444W} & Top-5 & 0 & 3 & 5 & 8\\
& Top-75\% & 0 & 3 & 5 & 8\\
& PSFEx & 0 & 4 & 5 & 9\\
\hline
\end{tabular}
\end{table}

\begin{table}[]
\caption{The number of host-galaxy detections for each final PSF}\label{tab:detnum}
\begin{tabular}{lcccccccccccc}
\hline\hline
 & \multicolumn{7}{c}{\# Detected filters} \\
\multirow{2}{*}{PSF} & \multicolumn{6}{c|}{COSMOS-Web} &  \multicolumn{5}{c}{PRIMER} \\
 & 0 & 1 & 2 & 3 & 4 & \multicolumn{1}{l|}{5} & 5 & 6 & 7 & 8 & 9 \\ \hline
Top-5 & 2 & 3 & 3 & 14 & 66 & \multicolumn{1}{l|}{14} & 0 & 0 & 2 & 0 & 3 \\ 
Top-75\% & 1 & 4 & 5 & 11 & 69 & \multicolumn{1}{l|}{12} & 0 & 2 & 0 & 0 & 3 \\ 
PSFEx & 2 & 3 & 3 & 15 & 68 & \multicolumn{1}{l|}{11} & 1 & 0 & 0 & 1 & 3 \\ 
\hline
\end{tabular}
\end{table}

\section{Detailed comparison of different PSF results}\label{app:different_PSF}
\begin{figure*}[ht!]\epsscale{1.15}
  \plotone{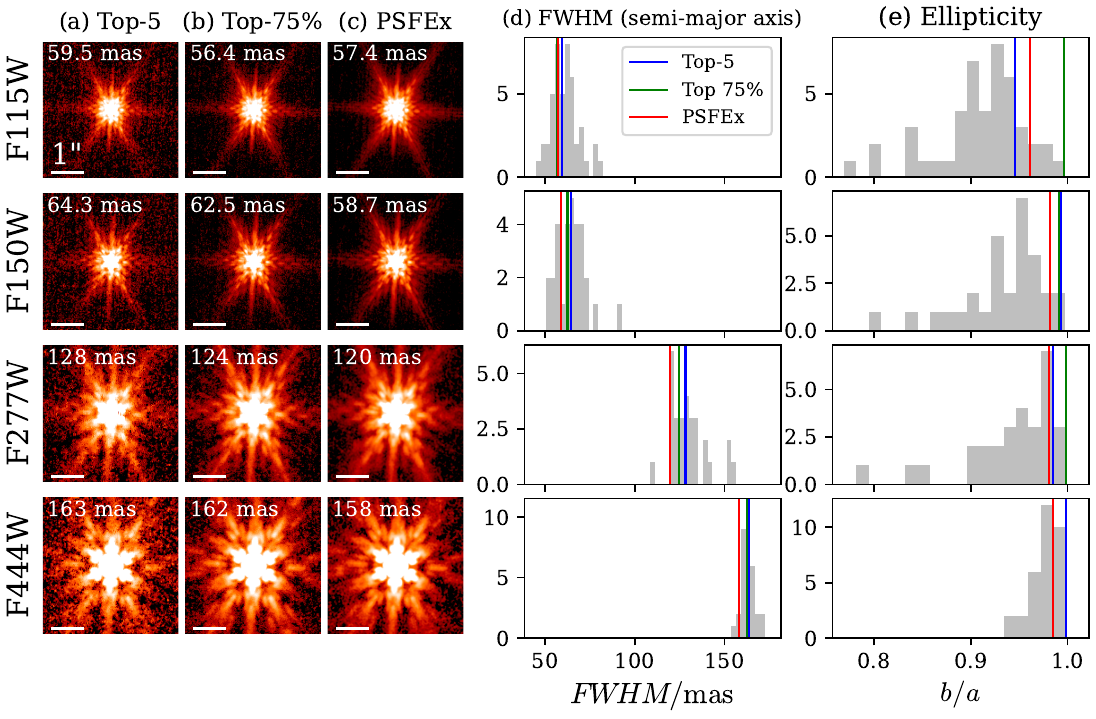}
  \caption{
  Comparison of different methods to evaluate empirical models of the PSF.
  Columns (a) to (c) show images of the top-5 stacked PSF, the top-75\% stacked PSF, and the {\tt PSFEx}-PSF for an example galaxy, CID-62 ($z=1.92$), in F115W, F150W, F277W, and F444W, from top to bottom.
  FWHM along the semi-major axis measured from the 2D Gaussian fits is shown in the top-left corner of each image.
  The white bars in the lower left indicate a scale of $1^{\prime\prime}$.
  Columns (d) and (e) show the distribution of the FWHM (semi-major axis) and ellipticity $b/a$.
  Gray histograms display the distribution of each single PSF in the library (Section~\ref{subsubsec:chisq_PSF}).
  The values for the top-5 stacked PSF, the top-75\% stacked PSF, {\tt PSFEx}-PSF are marked by the blue, green, and red vertical lines, respectively. 
  \label{fig:final_PSF}}
\end{figure*}

\begin{figure*}[ht!]\epsscale{1.15}
  \plotone{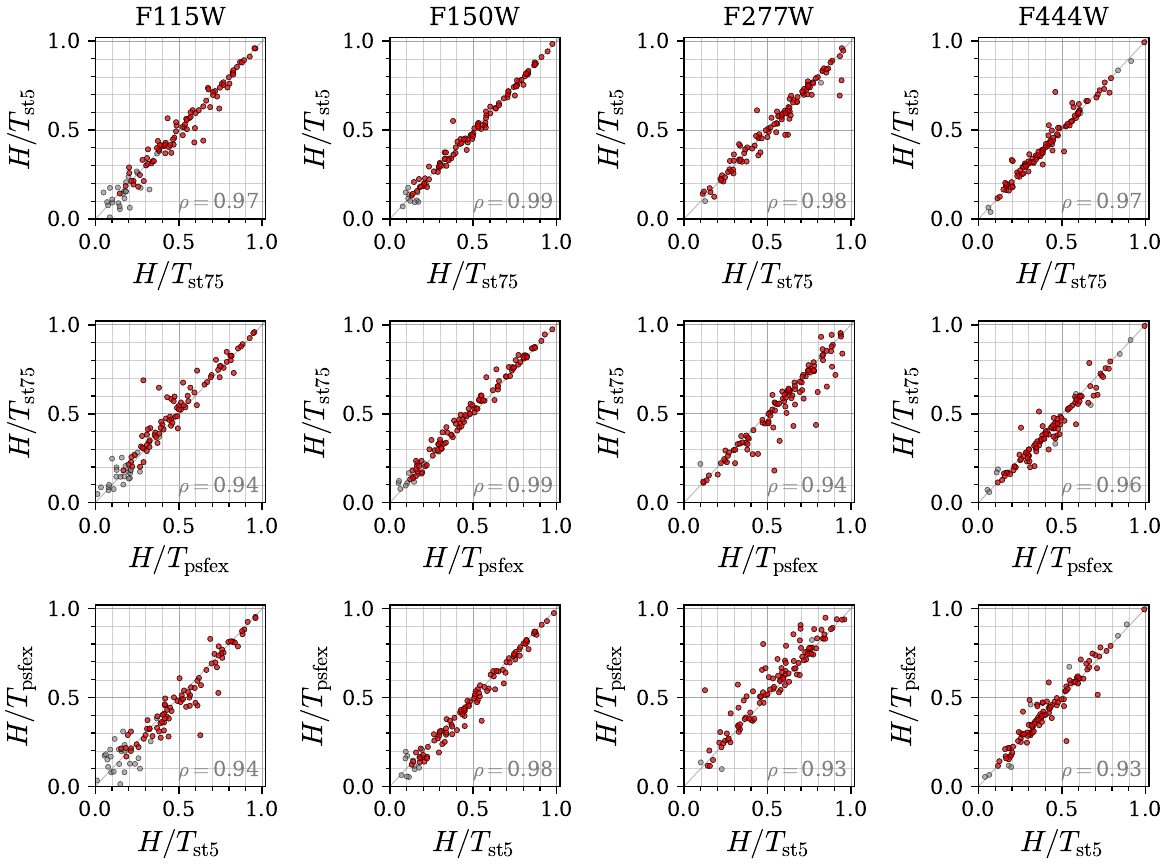}
  \caption{
   Comparison of the estimated $H/T$ in each filter using different PSFs (Columns~(a) to (d)).
  The gray and red colors correspond to whether they were detected in each filter.
  Spearman's correlation coefficient $\rho$ for each prior is shown in the lower right corner of each panel.
  High correlation coefficients and the distribution around $y=x$ (black dashed line) suggest consistent results among the fitting with different PSFs.
  \label{fig:ht_comp_full}}
\end{figure*}

As mentioned in Section~\ref{subsec:psf}, 2D decomposition is significantly influenced by the differences in PSFs.
In this study, we compare three different PSFs and discuss the PSF dependency of the results. 
Figure~\ref{fig:final_PSF} (images on the left; a--c) shows each final PSF image for the four NIRCam filters. We fit each final PSF and each single PSF in the PSF library with the 2D Gaussian model, and measure the FWHM along a semi-major axis and ellipticity $b/a$ (defined as $FWHM_{\rm minor}/FWHM_{\rm major}$).
Figures~\ref{fig:final_PSF} (d) and (e) show the distribution of FWHM and $b/a$ in each filter for an example AGN, CID-62, at $z_{\rm spec}\sim 1.92$.
As the number of stars used increases in the order of the top-5 stacked, the top-75\% stacked, and {\tt PSFEx}, we can see that the background noise is correspondingly lower.
Regarding the FWHM distribution, the top-5 and the top-75\% PSFs are consistent with the FWHM distributions of single PSFs within the PSF library.
{\tt PSFEx} have consistent FWHMs in F115W and F150W, and slightly smaller FWHMs in F277W and F444W, suggesting the possibility of bias between automatic PSF selection by {\tt PSFEx} and semi-automatic PSF selection by {\tt galight}.
In addition, each final PSF tends to have a higher $b/a$ than individual single PSFs.
Regardless of whether $\chi^2_\nu$-based stacking or empirical modeling is employed, considering that the final PSF is a more reasonable PSF reconstruction than a single PSF, this suggests a potential bias towards a more elliptical PSF when using a single PSF.
Next, Figure~\ref{fig:ht_comp_full} compares the estimated $H/T$ using each PSF and filter (the full-filter version of Figure~\ref{fig:comp_res_deco} (c)).
We can find strong positive correlations for all filters, with the distribution along the $y=x$ relation.
However, when comparing the distribution with $y=x$, it is evident that {\tt PSFEx} tends to estimate larger $H/T$ for F277W and F444W compared to top-5 and top-75\% PSFs.
As mentioned in Section~\ref{subsub:diff_final_psf}, this trend can be interpreted by considering the differences in FWHM for each filter and PSF.
As shown in Figure~\ref{fig:fwhm_comp}, {\tt PSFEx} tends to exhibit smaller PSFs in F277W and F444W than the top-5 and top-75\% PSFs.
\cite{Zhuang2023} suggest that different PSFs result in larger $H/T$ in the 2D decomposition analysis.
Therefore, it is conceivable that {\tt PSFEx} shows larger $H/T$ in F277W and F444W with the smaller PSFs than the top-5 and top-75\% PSFs.

\section{galight test with mock image}\label{app:galight_mock}
We generate mock images of AGN and their host galaxies and perform 2D decomposition on these images to verify whether {\tt galight} can accurately recover each parameter for the host galaxies.
We use {\tt lenstronomy} \citep{2018PDU....22..189B, 2021JOSS....6.3283B} to generate 2,000 NIRCam/F150W mock galaxy images. These are constructed as a superposition of a PSF and a smooth analytic \sersic model based on randomly sampled parameters: host-to-total flux ratio ($H/T$), \sersic index ($n$), effective radius ($r_e$), semi-major axis orientation ($\theta$), and ellipticity ($b/a$). 
The software tools {\tt lenstronomy} and {\tt galight} have different definitions of $r_e$, with the relation $r_{e,{\rm galight}} = \left(b/a\right)^{-0.5}\times r_{e,{\rm lenstronomy}}$.
In this section and consistent with the main part of this paper, all $r_e$ are unified under the definition in {\tt galight}.
The specified range for each parameter is $\log(H/T)$ in $\left[\sim-5, 0\right]$, $n$ in $\left[1, 6\right]$, $r_e$ in $\left[\sim0\farcs05,1\farcs6\right]$, $b/a$ in $\left[0.2,1.0\right]$, and $\theta$ in $\left[0,\pi\right]$ (in radian).
We assume the {\tt PSFEx} PSF (Section~\ref{subsubsec:modeling_PSF}) and that the center of host galaxy components are aligned with the center of PSF components.
Additionally, we add Gaussian noise equivalent to the real NIRCam/F150W COSMOS-Web images.
Using the generated mock images, we performed 2D decomposition with the low-$\chi^2_\nu$ top-5 PSF reconstruction strategy (Section~\ref{subsubsec:chisq_PSF}).
Since the {\tt PSFEx} PSF is assumed when generating the mock images, the results allow us to assess the performance in cases where the PSF may differ from the true PSF, similar to real data analysis.
Note that we cannot conclude which PSF reconstruction method is superior from this analysis. In this test, we do not consider the effects of more complex morphological features than single \sersic components, such as double \sersic (disc+bulge) components, position offset between AGN and host, substructures (bars, spirals, and clumps), and nearby or merging objects. 

\begin{figure*}[ht!]\epsscale{1.1}
  \plotone{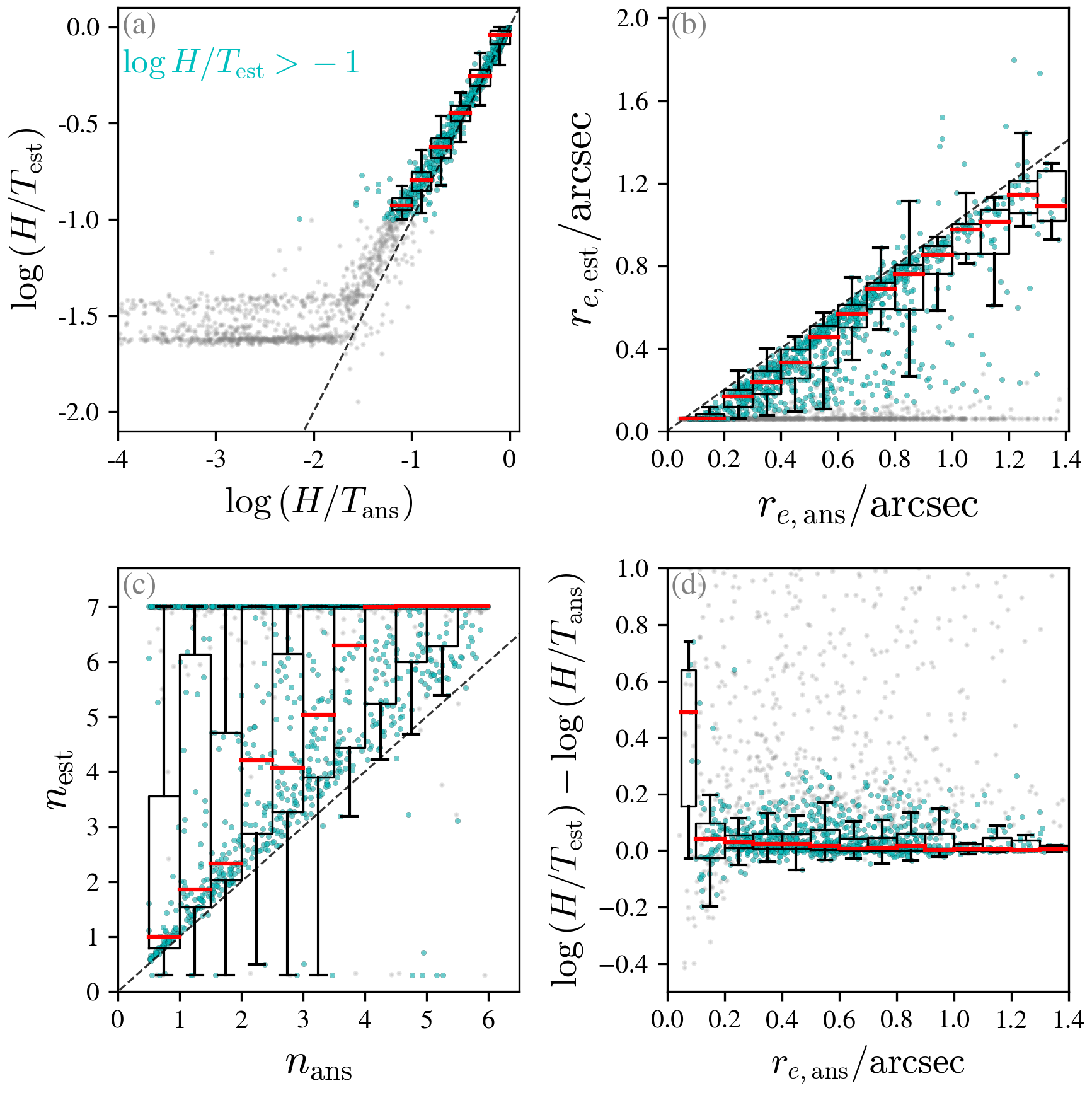}
  \caption{
  Comparison of input (X-axis) and estimated (Y-axis) morphological parameters of mock galaxies for the single Sérsic mock data.
  Columns (a) to (c) compare $\log\left(H/T\right)$, $r_e$, and $n$, respectively, and the black dashed line indicate $y=x$.
  The gray and cyan circles indicate the all mock galaxies and the galaxies with $\log\left(H/T_{\rm est}\right)>-1$ corresponding to the minimum $H/T_{\rm est}$ for our real galaxies (also see Figures~\ref{fig:comp_res_deco}, \ref{fig:ht_comp_full}).
  Only mock galaxies with $\log\left(H/T_{\rm est}\right) > -1$ are used in plotting box plots.
  As demonstrated, we can accurately reconstruct host galaxy photometry and morphology.
  \label{fig:est_ans}}
\end{figure*}

Figure~\ref{fig:est_ans} (a--c) compares the estimated major parameters ($H/T$, $r_e$, and $n$, respectively) with the input values.
First, focusing on $H/T$, we can find the distinct change in the trend around the input $\log\left(H/T\right)$ value $\log\left(H/T_{\rm ans}\right) \sim -2$.
For $\log\left(H/T_{\rm ans}\right) \lesssim -2$, the estimated value $\log\left(H/T_{\rm est}\right)$ remains relatively constant around $-1.5$, deviating significantly from the $y=x$ line.
This implies that detecting weak galaxy components with $\log\left(H/T_{\rm ans}\right) \lesssim-2$ is practically challenging using the 2D decomposition analysis with {\tt galight}.
On the other hand, for $\log\left(H/T_{\rm ans}\right) \gtrsim -2$, $\log\left(H/T_{\rm est}\right)$ exhibits a consistent trend along the $y=x$ relation, while slightly larger than the input values.
This trend is consistent with \cite{Zhuang2023}, who reported that different PSFs tend to overestimate $H/T$.
As shown in Figure~\ref{fig:param_dist}a, $\log\left(H/T_{\rm est}\right)$ for the real detected galaxies in this study are $\log\left(H/T_{\rm est}\right)\gtrsim-1$.
If we limit the mock results to $\log\left(H/T_{\rm est}\right)\gtrsim-1$, the difference between the estimation and input values (defined as estimation - input values) is $\Delta\log\left(H/T\right) \sim0.06\pm0.09~{\rm dex}$, suggesting sufficiently consistent estimation of $H/T$.
Next, examining $r_e$ and $n$, we find a consistent distribution along the $y=x$ relation, with an overall tendency to underestimate $r_e$ and overestimate $n$.
We can also see that some objects hit the $r_e$ lower limit ($r_e\sim0\farcs05$) and $n$ upper limit ($n\sim7$).
If we limit the analysis to the above realistic cases ($\log\left(H/T_{\rm est}\right)\gtrsim-1$), as shown in Figure~\ref{fig:est_ans}b, $r_e$ shows consistent results with $\Delta r = -0.14\pm 0.17$ and less hitting to the $r_e$ lower limit.
However, for $n$, many objects still exhibit the $n$ upper limit even with $\log\left(H/T_{\rm est}\right)\gtrsim-1$. 
If we additionally cut the upper limit object with $n<6.8$, the difference between the estimation and the input values is $\Delta n \sim 1.4\pm 1.6$, suggesting a consistent estimation with medium scatter.
We confirm that larger-$n$ objects tend to have larger $\Delta n$, indicating the increasing difficulty of accurately reconstructing $n$ when they have larger $n$.
Examining the residual $H/T$ (defined as $\log\left(H/T_{\rm est}\right)-\log\left(H/T_{\rm ans}\right)$) as a function of $r_{e, {\rm ans}}$ (Figure~\ref{fig:est_ans}d), we confirm that, with $\log\left(H/T_{\rm est}\right)>-1$, the values of $H/T$ can be recovered independently over a wide $r_e$ range of $\sim$0\farcs1-1\farcs2, effectively covering the typical $r_e$ of the sample (Figure~\ref{fig:param_dist}b).

\begin{figure*}[ht!]\epsscale{1.15}
  \plotone{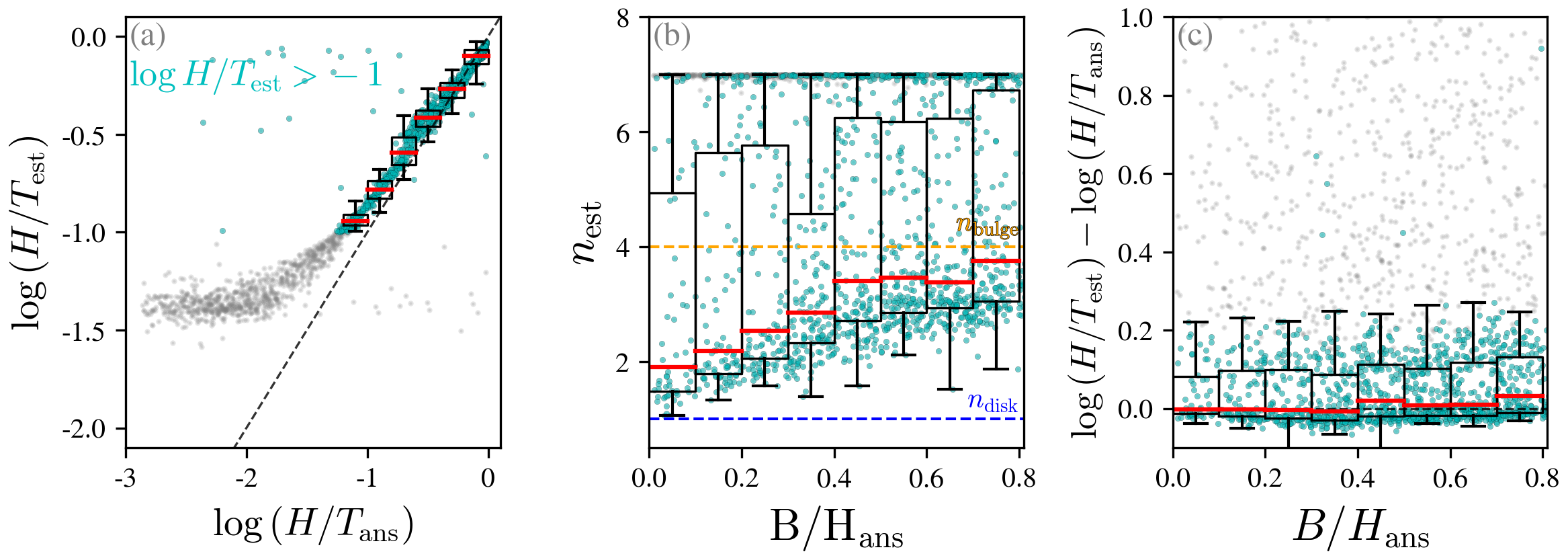}
  \caption{
  (a) Comparison of the input and estimated $\log\left(H/T\right)$ for the double Sérsic mock data.
  (b) The estimated $n$ as a function of the input $B/H$. Blue and orange dashed lines indicate $n_{\rm disk}$ and $n_{\rm bulge}$.
  (c) The $\log\left(H/T\right)$ residual as a function of the input $B/H$.
  In every panel, the gray and cyan circles indicate the all mock galaxies and the galaxies with $\log\left(H/T_{\rm est}\right)>-1$ corresponding to the minimum $H/T_{\rm est}$ for our real galaxies (also see Figures~\ref{fig:comp_res_deco}, \ref{fig:ht_comp_full}).
  Only mock galaxies with $\log\left(H/T_{\rm est}\right) > -1$ are used for the box plots.
  As shown, we can accurately reconstruct $H/T$ flux ratio independent of bulge dominance $B/H$.
  \label{fig:est_ans_ds}}
\end{figure*}

In this paper, we assume the PSF + single Sérsic model to fit the galaxy images.
As discussed in section~\ref{subsec:decomposition}, single Sérsic model is a first-order approximation and cannot fully describe galaxy morphologies.
For example, substructures such as bars and spiral arms are not considered, and both disk and bulge components cannot be described with a single Sérsic component.
Even so, we expand the above mock tests using a PSF + double Sérsic model to address the effect of a bulge component on the PSF + single Sérsic model fitting.
One of the double Sérsic components corresponds to a disk, and we assume $r_{e,{\rm disk}}=0\farcs35$ and $n_{\rm disk}=1.0$.
The other Sérsic component corresponds to a bulge, and we assume $n_{\rm bulge}=4.0$.
$r_{e,{\rm bulge}}$, $H/T$, bulge-to-host flux ratio $B/H$, $\theta_{\rm disk}$, $\theta_{\rm bulge}$, $b/a_{\rm disk}$, and $b/a_{\rm bulge}$ is randomly sampled.
The specified range for each parameter is $\log(H/T)$ in $\left[\sim-3, 0\right]$, $r_{e,{\rm bulge}}$ in $\left[\sim 0\farcs06, 0\farcs2 \right]$, $b/a_{\rm disk}$ in $\left[0.6,1.0\right]$, $b/a_{\rm bulge}$ in $\left[0.8, 1.0\right]$, $\theta_{\rm disk}$ in $\left[0,\pi\right]$ (in radian), and $\theta_{\rm bulge}$ in $\left[0,\pi\right]$ (in radian).
Using these models and randomly sampled parameters, we make mock data by following the same method with the above single Sérsic mock data.
Then, we performed 2D decomposition with the low-$\chi^2_\nu$ top-5 PSF reconstruction strategy (Section~\ref{subsubsec:chisq_PSF}), and the results are summarized in figure~\ref{fig:est_ans_ds}.
Figure~\ref{fig:est_ans_ds} compares the estimated $H/T$ with the input values.
Due to the inclusion of the bulge model, the deviation from the $y=x$ line around $\log\left(H/T\right)\sim-2$ is larger than for the single Sérsic mock data (Figure~\ref{fig:est_ans}a). 
However, similar to the results for the single Sérsic mock data, $H/T$ is well reconstructed for $\log\left(H/T_{\rm est}\right) > -1$.
Figure~\ref{fig:est_ans_ds}b and c indicate the $B/H$ dependence of $n_{\rm est}$ and $\log\left(H/T\right)$ residual.
From figure~\ref{fig:est_ans_ds}b, as $B/H_{\rm ans}$ increases and the system becomes more bulge-dominated, the estimated $n$ approaches values from 1 ($=n_{\rm disk}$) to 4 ($=n_{\rm bulge}$), indicating that the single Sérsic model fits the bulge for a bulge-dominated system.
However, from figure~\ref{fig:est_ans_ds}c, $H/T$ is well reconstructed independently on the $B/H$ in the range of $B/H\sim0-0.6$.
Therefore, we conclude that we can reconstruct $H/T$, i.e., host photometry, well for systems where the bulge constitutes at most $\sim$50\% of the total host galaxy flux.

\bibliography{export}{}
\bibliographystyle{aasjournal}

\end{document}